\documentclass[twocolumn,twocolappendix,trackchanges]{aastex631}

\usepackage{amsmath}
\usepackage{enumitem}

\begin{document}

\title{How to Detect an Astrophysical Nanohertz Gravitational-Wave Background}

\correspondingauthor{Bence B\'ecsy}
\email{becsyb@oregonstate.edu}

\author[0000-0003-0909-5563]{Bence B\'ecsy}
\affiliation{Department of Physics, Oregon State University, Corvallis, OR 97331, USA}

\author[0000-0002-7435-0869]{Neil J.~Cornish}
\affiliation{eXtreme Gravity Institute, Department of Physics, Montana State University, Bozeman, MT 59717, USA}%

\author[0000-0002-2689-0190]{Patrick M.~Meyers}
\affiliation{Division of Physics, Mathematics, and Astronomy, California Institute of Technology, Pasadena, CA 91125, USA}

\author[0000-0002-6625-6450]{Luke Zoltan Kelley}
\affiliation{Department of Astronomy, University of California at Berkeley, Berkeley, CA 94720, USA}%

\author[0000-0001-5134-3925]{Gabriella Agazie}
\affiliation{Center for Gravitation, Cosmology and Astrophysics, Department of Physics, University of Wisconsin-Milwaukee,\\ P.O. Box 413, Milwaukee, WI 53201, USA}

\author[0000-0002-8935-9882]{Akash Anumarlapudi}
\affiliation{Center for Gravitation, Cosmology and Astrophysics, Department of Physics, University of Wisconsin-Milwaukee,\\ P.O. Box 413, Milwaukee, WI 53201, USA}

\author[0000-0003-0638-3340]{Anne M. Archibald}
\affiliation{Newcastle University, NE1 7RU, UK}

\author{Zaven Arzoumanian}
\affiliation{X-Ray Astrophysics Laboratory, NASA Goddard Space Flight Center, Code 662, Greenbelt, MD 20771, USA}

\author[0000-0003-2745-753X]{Paul T. Baker}
\affiliation{Department of Physics and Astronomy, Widener University, One University Place, Chester, PA 19013, USA}

\author[0000-0002-2183-1087]{Laura Blecha}
\affiliation{Physics Department, University of Florida, Gainesville, FL 32611, USA}

\author[0000-0001-6341-7178]{Adam Brazier}
\affiliation{Cornell Center for Astrophysics and Planetary Science and Department of Astronomy, Cornell University, Ithaca, NY 14853, USA}
\affiliation{Cornell Center for Advanced Computing, Cornell University, Ithaca, NY 14853, USA}

\author[0000-0003-3053-6538]{Paul R. Brook}
\affiliation{Institute for Gravitational Wave Astronomy and School of Physics and Astronomy, University of Birmingham, Edgbaston, Birmingham B15 2TT, UK}

\author[0000-0003-4052-7838]{Sarah Burke-Spolaor}
\altaffiliation{Sloan Fellow}
\affiliation{Department of Physics and Astronomy, West Virginia University, P.O. Box 6315, Morgantown, WV 26506, USA}
\affiliation{Center for Gravitational Waves and Cosmology, West Virginia University, Chestnut Ridge Research Building, Morgantown, WV 26505, USA}

\author[0000-0002-5557-4007]{J. Andrew Casey-Clyde}
\affiliation{Department of Physics, University of Connecticut, 196 Auditorium Road, U-3046, Storrs, CT 06269-3046, USA}

\author[0000-0003-3579-2522]{Maria Charisi}
\affiliation{Department of Physics and Astronomy, Vanderbilt University, 2301 Vanderbilt Place, Nashville, TN 37235, USA}

\author[0000-0002-2878-1502]{Shami Chatterjee}
\affiliation{Cornell Center for Astrophysics and Planetary Science and Department of Astronomy, Cornell University, Ithaca, NY 14853, USA}

\author{Katerina Chatziioannou}
\affiliation{Division of Physics, Mathematics, and Astronomy, California Institute of Technology, Pasadena, CA 91125, USA}

\author[0000-0001-7587-5483]{Tyler Cohen}
\affiliation{Department of Physics, New Mexico Institute of Mining and Technology, 801 Leroy Place, Socorro, NM 87801, USA}

\author[0000-0002-4049-1882]{James M. Cordes}
\affiliation{Cornell Center for Astrophysics and Planetary Science and Department of Astronomy, Cornell University, Ithaca, NY 14853, USA}

\author[0000-0002-2578-0360]{Fronefield Crawford}
\affiliation{Department of Physics and Astronomy, Franklin \& Marshall College, P.O. Box 3003, Lancaster, PA 17604, USA}

\author[0000-0002-6039-692X]{H. Thankful Cromartie}
\altaffiliation{NASA Hubble Fellowship: Einstein Postdoctoral Fellow}
\affiliation{Cornell Center for Astrophysics and Planetary Science and Department of Astronomy, Cornell University, Ithaca, NY 14853, USA}

\author[0000-0002-1529-5169]{Kathryn Crowter}
\affiliation{Department of Physics and Astronomy, University of British Columbia, 6224 Agricultural Road, Vancouver, BC V6T 1Z1, Canada}

\author[0000-0002-2185-1790]{Megan E. DeCesar}
\affiliation{George Mason University, resident at the Naval Research Laboratory, Washington, DC 20375, USA}

\author[0000-0002-6664-965X]{Paul B. Demorest}
\affiliation{National Radio Astronomy Observatory, 1003 Lopezville Rd., Socorro, NM 87801, USA}

\author[0000-0001-8885-6388]{Timothy Dolch}
\affiliation{Department of Physics, Hillsdale College, 33 E. College Street, Hillsdale, MI 49242, USA}
\affiliation{Eureka Scientific, 2452 Delmer Street, Suite 100, Oakland, CA 94602-3017, USA}

\author[0000-0001-7828-7708]{Elizabeth C. Ferrara}
\affiliation{Department of Astronomy, University of Maryland, College Park, MD 20742}
\affiliation{Center for Research and Exploration in Space Science and Technology, NASA/GSFC, Greenbelt, MD 20771}
\affiliation{NASA Goddard Space Flight Center, Greenbelt, MD 20771, USA}

\author[0000-0001-5645-5336]{William Fiore}
\affiliation{Department of Physics and Astronomy, West Virginia University, P.O. Box 6315, Morgantown, WV 26506, USA}
\affiliation{Center for Gravitational Waves and Cosmology, West Virginia University, Chestnut Ridge Research Building, Morgantown, WV 26505, USA}

\author[0000-0001-8384-5049]{Emmanuel Fonseca}
\affiliation{Department of Physics and Astronomy, West Virginia University, P.O. Box 6315, Morgantown, WV 26506, USA}
\affiliation{Center for Gravitational Waves and Cosmology, West Virginia University, Chestnut Ridge Research Building, Morgantown, WV 26505, USA}

\author[0000-0001-7624-4616]{Gabriel E. Freedman}
\affiliation{Center for Gravitation, Cosmology and Astrophysics, Department of Physics, University of Wisconsin-Milwaukee,\\ P.O. Box 413, Milwaukee, WI 53201, USA}

\author[0000-0001-6166-9646]{Nate Garver-Daniels}
\affiliation{Department of Physics and Astronomy, West Virginia University, P.O. Box 6315, Morgantown, WV 26506, USA}
\affiliation{Center for Gravitational Waves and Cosmology, West Virginia University, Chestnut Ridge Research Building, Morgantown, WV 26505, USA}

\author[0000-0001-8158-683X]{Peter A. Gentile}
\affiliation{Department of Physics and Astronomy, West Virginia University, P.O. Box 6315, Morgantown, WV 26506, USA}
\affiliation{Center for Gravitational Waves and Cosmology, West Virginia University, Chestnut Ridge Research Building, Morgantown, WV 26505, USA}

\author[0000-0003-4090-9780]{Joseph Glaser}
\affiliation{Department of Physics and Astronomy, West Virginia University, P.O. Box 6315, Morgantown, WV 26506, USA}
\affiliation{Center for Gravitational Waves and Cosmology, West Virginia University, Chestnut Ridge Research Building, Morgantown, WV 26505, USA}

\author[0000-0003-1884-348X]{Deborah C. Good}
\affiliation{Department of Physics, University of Connecticut, 196 Auditorium Road, U-3046, Storrs, CT 06269-3046, USA}
\affiliation{Center for Computational Astrophysics, Flatiron Institute, 162 5th Avenue, New York, NY 10010, USA}

\author[0000-0002-1146-0198]{Kayhan G\"{u}ltekin}
\affiliation{Department of Astronomy and Astrophysics, University of Michigan, Ann Arbor, MI 48109, USA}

\author[0000-0003-2742-3321]{Jeffrey S. Hazboun}
\affiliation{Department of Physics, Oregon State University, Corvallis, OR 97331, USA}

\author[0000-0002-9152-0719]{Sophie Hourihane}
\affiliation{Division of Physics, Mathematics, and Astronomy, California Institute of Technology, Pasadena, CA 91125, USA}

\author[0000-0003-1082-2342]{Ross J. Jennings}
\altaffiliation{NANOGrav Physics Frontiers Center Postdoctoral Fellow}
\affiliation{Department of Physics and Astronomy, West Virginia University, P.O. Box 6315, Morgantown, WV 26506, USA}
\affiliation{Center for Gravitational Waves and Cosmology, West Virginia University, Chestnut Ridge Research Building, Morgantown, WV 26505, USA}

\author[0000-0002-7445-8423]{Aaron D. Johnson}
\affiliation{Center for Gravitation, Cosmology and Astrophysics, Department of Physics, University of Wisconsin-Milwaukee,\\ P.O. Box 413, Milwaukee, WI 53201, USA}
\affiliation{Division of Physics, Mathematics, and Astronomy, California Institute of Technology, Pasadena, CA 91125, USA}

\author[0000-0001-6607-3710]{Megan L. Jones}
\affiliation{Center for Gravitation, Cosmology and Astrophysics, Department of Physics, University of Wisconsin-Milwaukee,\\ P.O. Box 413, Milwaukee, WI 53201, USA}

\author[0000-0002-3654-980X]{Andrew R. Kaiser}
\affiliation{Department of Physics and Astronomy, West Virginia University, P.O. Box 6315, Morgantown, WV 26506, USA}
\affiliation{Center for Gravitational Waves and Cosmology, West Virginia University, Chestnut Ridge Research Building, Morgantown, WV 26505, USA}

\author[0000-0001-6295-2881]{David L. Kaplan}
\affiliation{Center for Gravitation, Cosmology and Astrophysics, Department of Physics, University of Wisconsin-Milwaukee,\\ P.O. Box 413, Milwaukee, WI 53201, USA}

\author[0000-0002-0893-4073]{Matthew Kerr}
\affiliation{Space Science Division, Naval Research Laboratory, Washington, DC 20375-5352, USA}

\author[0000-0003-0123-7600]{Joey S. Key}
\affiliation{University of Washington Bothell, 18115 Campus Way NE, Bothell, WA 98011, USA}

\author[0000-0002-9197-7604]{Nima Laal}
\affiliation{Department of Physics, Oregon State University, Corvallis, OR 97331, USA}

\author[0000-0003-0721-651X]{Michael T. Lam}
\affiliation{SETI Institute, 339 N Bernardo Ave Suite 200, Mountain View, CA 94043, USA}
\affiliation{School of Physics and Astronomy, Rochester Institute of Technology, Rochester, NY 14623, USA}
\affiliation{Laboratory for Multiwavelength Astrophysics, Rochester Institute of Technology, Rochester, NY 14623, USA}

\author[0000-0003-1096-4156]{William G. Lamb}
\affiliation{Department of Physics and Astronomy, Vanderbilt University, 2301 Vanderbilt Place, Nashville, TN 37235, USA}

\author{T. Joseph W. Lazio}
\affiliation{Jet Propulsion Laboratory, California Institute of Technology, 4800 Oak Grove Drive, Pasadena, CA 91109, USA}

\author[0000-0003-0771-6581]{Natalia Lewandowska}
\affiliation{Department of Physics, State University of New York at Oswego, Oswego, NY, 13126, USA}

\author[0000-0002-9574-578X]{Tyson B. Littenberg}
\affiliation{NASA Marshall Space Flight Center, Huntsville, AL 35812, USA}

\author[0000-0001-5766-4287]{Tingting Liu}
\affiliation{Department of Physics and Astronomy, West Virginia University, P.O. Box 6315, Morgantown, WV 26506, USA}
\affiliation{Center for Gravitational Waves and Cosmology, West Virginia University, Chestnut Ridge Research Building, Morgantown, WV 26505, USA}

\author[0000-0003-1301-966X]{Duncan R. Lorimer}
\affiliation{Department of Physics and Astronomy, West Virginia University, P.O. Box 6315, Morgantown, WV 26506, USA}
\affiliation{Center for Gravitational Waves and Cosmology, West Virginia University, Chestnut Ridge Research Building, Morgantown, WV 26505, USA}

\author[0000-0001-5373-5914]{Jing Luo}
\altaffiliation{Deceased}
\affiliation{Department of Astronomy \& Astrophysics, University of Toronto, 50 Saint George Street, Toronto, ON M5S 3H4, Canada}

\author[0000-0001-5229-7430]{Ryan S. Lynch}
\affiliation{Green Bank Observatory, P.O. Box 2, Green Bank, WV 24944, USA}

\author[0000-0002-4430-102X]{Chung-Pei Ma}
\affiliation{Department of Astronomy, University of California, Berkeley, 501 Campbell Hall \#3411, Berkeley, CA 94720, USA}
\affiliation{Department of Physics, University of California, Berkeley, CA 94720, USA}

\author[0000-0003-2285-0404]{Dustin R. Madison}
\affiliation{Department of Physics, University of the Pacific, 3601 Pacific Avenue, Stockton, CA 95211, USA}

\author[0000-0001-5481-7559]{Alexander McEwen}
\affiliation{Center for Gravitation, Cosmology and Astrophysics, Department of Physics, University of Wisconsin-Milwaukee,\\ P.O. Box 413, Milwaukee, WI 53201, USA}

\author[0000-0002-2885-8485]{James W. McKee}
\affiliation{E.A. Milne Centre for Astrophysics, University of Hull, Cottingham Road, Kingston-upon-Hull, HU6 7RX, UK}
\affiliation{Centre of Excellence for Data Science, Artificial Intelligence and Modelling (DAIM), University of Hull, Cottingham Road, Kingston-upon-Hull, HU6 7RX, UK}

\author[0000-0001-7697-7422]{Maura A. McLaughlin}
\affiliation{Department of Physics and Astronomy, West Virginia University, P.O. Box 6315, Morgantown, WV 26506, USA}
\affiliation{Center for Gravitational Waves and Cosmology, West Virginia University, Chestnut Ridge Research Building, Morgantown, WV 26505, USA}

\author[0000-0002-4642-1260]{Natasha McMann}
\affiliation{Department of Physics and Astronomy, Vanderbilt University, 2301 Vanderbilt Place, Nashville, TN 37235, USA}

\author[0000-0001-8845-1225]{Bradley W. Meyers}
\affiliation{Department of Physics and Astronomy, University of British Columbia, 6224 Agricultural Road, Vancouver, BC V6T 1Z1, Canada}
\affiliation{International Centre for Radio Astronomy Research, Curtin University, Bentley, WA 6102, Australia}

\author[0000-0002-4307-1322]{Chiara M. F. Mingarelli}
\affiliation{Department of Physics, Yale University, New Haven, CT 06520, USA}

\author[0000-0003-2898-5844]{Andrea Mitridate}
\affiliation{Deutsches Elektronen-Synchrotron DESY, Notkestr. 85, 22607 Hamburg, Germany}

\author[0000-0002-3616-5160]{Cherry Ng}
\affiliation{Dunlap Institute for Astronomy and Astrophysics, University of Toronto, 50 St. George St., Toronto, ON M5S 3H4, Canada}

\author[0000-0002-6709-2566]{David J. Nice}
\affiliation{Department of Physics, Lafayette College, Easton, PA 18042, USA}

\author[0000-0002-4941-5333]{Stella Koch Ocker}
\affiliation{Division of Physics, Mathematics, and Astronomy, California Institute of Technology, Pasadena, CA 91125, USA}
\affiliation{The Observatories of the Carnegie Institution for Science, Pasadena, CA 91101, USA}

\author[0000-0002-2027-3714]{Ken D. Olum}
\affiliation{Institute of Cosmology, Department of Physics and Astronomy, Tufts University, Medford, MA 02155, USA}

\author[0000-0001-5465-2889]{Timothy T. Pennucci}
\affiliation{Institute of Physics and Astronomy, E\"{o}tv\"{o}s Lor\'{a}nd University, P\'{a}zm\'{a}ny P. s. 1/A, 1117 Budapest, Hungary}

\author[0000-0002-8509-5947]{Benetge B. P. Perera}
\affiliation{Arecibo Observatory, HC3 Box 53995, Arecibo, PR 00612, USA}

\author[0000-0002-8826-1285]{Nihan S. Pol}
\affiliation{Department of Physics and Astronomy, Vanderbilt University, 2301 Vanderbilt Place, Nashville, TN 37235, USA}

\author[0000-0002-2074-4360]{Henri A. Radovan}
\affiliation{Department of Physics, University of Puerto Rico, Mayag\"{u}ez, PR 00681, USA}

\author[0000-0001-5799-9714]{Scott M. Ransom}
\affiliation{National Radio Astronomy Observatory, 520 Edgemont Road, Charlottesville, VA 22903, USA}

\author[0000-0002-5297-5278]{Paul S. Ray}
\affiliation{Space Science Division, Naval Research Laboratory, Washington, DC 20375-5352, USA}

\author[0000-0003-4915-3246]{Joseph D. Romano}
\affiliation{Department of Physics, Texas Tech University, Box 41051, Lubbock, TX 79409, USA}

\author[0009-0006-5476-3603]{Shashwat C. Sardesai}
\affiliation{Center for Gravitation, Cosmology and Astrophysics, Department of Physics, University of Wisconsin-Milwaukee,\\ P.O. Box 413, Milwaukee, WI 53201, USA}

\author[0000-0003-4391-936X]{Ann Schmiedekamp}
\affiliation{Department of Physics, Penn State Abington, Abington, PA 19001, USA}

\author[0000-0002-1283-2184]{Carl Schmiedekamp}
\affiliation{Department of Physics, Penn State Abington, Abington, PA 19001, USA}

\author[0000-0003-2807-6472]{Kai Schmitz}
\affiliation{Institute for Theoretical Physics, University of M\"{u}nster, 48149 M\"{u}nster, Germany}

\author[0000-0002-7283-1124]{Brent J. Shapiro-Albert}
\affiliation{Department of Physics and Astronomy, West Virginia University, P.O. Box 6315, Morgantown, WV 26506, USA}
\affiliation{Center for Gravitational Waves and Cosmology, West Virginia University, Chestnut Ridge Research Building, Morgantown, WV 26505, USA}
\affiliation{Giant Army, 915A 17th Ave, Seattle WA 98122}

\author[0000-0002-7778-2990]{Xavier Siemens}
\affiliation{Department of Physics, Oregon State University, Corvallis, OR 97331, USA}
\affiliation{Center for Gravitation, Cosmology and Astrophysics, Department of Physics, University of Wisconsin-Milwaukee,\\ P.O. Box 413, Milwaukee, WI 53201, USA}

\author[0000-0003-1407-6607]{Joseph Simon}
\altaffiliation{NSF Astronomy and Astrophysics Postdoctoral Fellow}
\affiliation{Department of Astrophysical and Planetary Sciences, University of Colorado, Boulder, CO 80309, USA}

\author[0000-0002-1530-9778]{Magdalena S. Siwek}
\affiliation{Center for Astrophysics, Harvard University, 60 Garden St, Cambridge, MA 02138}

\author[0000-0002-5176-2924]{Sophia V. Sosa Fiscella}
\affiliation{School of Physics and Astronomy, Rochester Institute of Technology, Rochester, NY 14623, USA}
\affiliation{Laboratory for Multiwavelength Astrophysics, Rochester Institute of Technology, Rochester, NY 14623, USA}

\author[0000-0001-9784-8670]{Ingrid H. Stairs}
\affiliation{Department of Physics and Astronomy, University of British Columbia, 6224 Agricultural Road, Vancouver, BC V6T 1Z1, Canada}

\author[0000-0002-1797-3277]{Daniel R. Stinebring}
\affiliation{Department of Physics and Astronomy, Oberlin College, Oberlin, OH 44074, USA}

\author[0000-0002-7261-594X]{Kevin Stovall}
\affiliation{National Radio Astronomy Observatory, 1003 Lopezville Rd., Socorro, NM 87801, USA}

\author[0000-0002-2820-0931]{Abhimanyu Susobhanan}
\affiliation{Center for Gravitation, Cosmology and Astrophysics, Department of Physics, University of Wisconsin-Milwaukee,\\ P.O. Box 413, Milwaukee, WI 53201, USA}

\author[0000-0002-1075-3837]{Joseph K. Swiggum}
\altaffiliation{NANOGrav Physics Frontiers Center Postdoctoral Fellow}
\affiliation{Department of Physics, Lafayette College, Easton, PA 18042, USA}

\author[0000-0003-0264-1453]{Stephen R. Taylor}
\affiliation{Department of Physics and Astronomy, Vanderbilt University, 2301 Vanderbilt Place, Nashville, TN 37235, USA}

\author[0000-0002-2451-7288]{Jacob E. Turner}
\affiliation{Green Bank Observatory, P.O. Box 2, Green Bank, WV 24944, USA}

\author[0000-0001-8800-0192]{Caner Unal}
\affiliation{Department of Physics, Middle East Technical University, 06531 Ankara, Turkey}
\affiliation{Department of Physics, Ben-Gurion University of the Negev, Be'er Sheva 84105, Israel}
\affiliation{Feza Gursey Institute, Bogazici University, Kandilli, 34684, Istanbul, Turkey}

\author[0000-0002-4162-0033]{Michele Vallisneri}
\affiliation{Jet Propulsion Laboratory, California Institute of Technology, 4800 Oak Grove Drive, Pasadena, CA 91109, USA}
\affiliation{Division of Physics, Mathematics, and Astronomy, California Institute of Technology, Pasadena, CA 91125, USA}

\author[0000-0002-6428-2620]{Rutger van~Haasteren}
\affiliation{Max-Planck-Institut f\"{u}r Gravitationsphysik (Albert-Einstein-Institut), Callinstrasse 38, D-30167, Hannover, Germany}

\author[0000-0003-4700-9072]{Sarah J. Vigeland}
\affiliation{Center for Gravitation, Cosmology and Astrophysics, Department of Physics, University of Wisconsin-Milwaukee,\\ P.O. Box 413, Milwaukee, WI 53201, USA}

\author[0000-0001-9678-0299]{Haley M. Wahl}
\affiliation{Department of Physics and Astronomy, West Virginia University, P.O. Box 6315, Morgantown, WV 26506, USA}
\affiliation{Center for Gravitational Waves and Cosmology, West Virginia University, Chestnut Ridge Research Building, Morgantown, WV 26505, USA}

\author[0000-0002-6020-9274]{Caitlin A. Witt}
\affiliation{Center for Interdisciplinary Exploration and Research in Astrophysics (CIERA), Northwestern University, Evanston, IL 60208}
\affiliation{Adler Planetarium, 1300 S. DuSable Lake Shore Dr., Chicago, IL 60605, USA}

\author[0000-0002-0883-0688]{Olivia Young}
\affiliation{School of Physics and Astronomy, Rochester Institute of Technology, Rochester, NY 14623, USA}
\affiliation{Laboratory for Multiwavelength Astrophysics, Rochester Institute of Technology, Rochester, NY 14623, USA}


\begin{abstract}
Analyses of pulsar timing data have provided evidence for a stochastic gravitational wave background in the nHz frequency band. The most plausible source of such a background is the superposition of signals from millions of supermassive black hole binaries. The standard statistical techniques used to search for such a background and assess its significance make several simplifying assumptions, namely: i) Gaussianity; ii) isotropy; and most often iii) a power-law spectrum. However, a stochastic background from a finite collection of binaries does not exactly satisfy any of these assumptions. To understand the effect of these assumptions, we test standard analysis techniques on a large collection of realistic simulated datasets. The dataset length, observing schedule, and noise levels were chosen to emulate the NANOGrav 15-year dataset.  Simulated signals from millions of binaries drawn from models based on the Illustris cosmological hydrodynamical simulation were added to the data. We find that the standard statistical methods perform remarkably well on these simulated datasets, despite their fundamental assumptions not being strictly met. They are able to achieve a confident detection of the background. However, even for a fixed set of astrophysical parameters, different realizations of the universe result in a large variance in the significance and recovered parameters of the background. We also find that the presence of loud individual binaries can bias the spectral recovery of the background if we do not account for them.
\end{abstract}

\section{\label{sec:intro}Introduction}
Pulsar timing arrays (PTAs) monitor millisecond pulsars over a timespan of decades. The times-of-arrival (TOAs) of these radio pulses are sensitive to various physical processes, including variations in the proper distance between the pulsars and the observer due to passing gravitational waves (GWs). PTAs recently reported evidence for a nanohertz stochastic GW background (GWB) in their most recent datasets \citep{nanograv_15yr_gwb, epta_dr2_gwb, ppta_dr3_gwb, cpta_gwb}, and their results were shown to be consistent with each other \citep{3p_plus_comparison}. The most plausible source of such a background is a large collection of supermassive black hole binaries (SMBHBs, \citealt{nanograv_15yr_astro}), but other more exotic phenomena can also explain the signal \citep{nanograv_15yr_new_physics}.

Several methods have been developed to search for a GWB in PTA datasets. While the GWB is expected to first appear as a common uncorrelated red noise process (CURN), the definitive signature of a GWB is considered to be the characteristic Hellings-Downs (HD, \citealt{HD}) angular correlation pattern between pulsars. The most notable frequentist method is the so called optimal statistic (OS, \citealt{OS}), which is an unbiased maximum-likelihood estimator of the GWB amplitude based on the cross-correlations between pulsars and the theoretical correlation pattern. It also allows one to define a signal-to-noise ratio (S/N), which is used as a detection statistic. Bayesian methods, on the other hand, marginalize over the parameters of a Fourier basis Gaussian process that describes the GWB \citep{lentati_et_al_2013, rutger_michele_2014}. This can be constructed either with (HD model) or without HD correlations between pulsars (CURN model). One can calculate the Bayes factor (BF) between these two models, which is the most important quantity in quantifying the level of Bayesian evidence for a GWB.

As with any statistical data analysis, one needs to make various assumptions when calculating either the OS S/N or the HD vs.~CURN BF. Most commonly one assumes that the GWB is: i) Gaussian; ii) isotropic; iii) described by a power-law spectrum\footnote{This also means we implicitly assume that the signal is stationary.}. However, if the GWB is produced by a finite number of SMBHBs, none of these assumptions hold exactly (see e.g.~\citealt{Becsy:2022pnr}). Methods have been developed that relax some of these assumptions. Assumption iii) is the one most often relaxed either by modeling the spectrum as a two-component so called broken power-law model or allowing the amplitudes of different frequency components to vary freely (see e.g.~\citealt{Sampson:2015ada, Steve_astro_spectral_modeling, nanograv_15yr_gwb, nanograv_15yr_detchar, Pat_post_Pred_check, William_spectral_refitting, free_spectrum_os}). There are also specific searches looking for an anisotropic GWB \citep{Chiara_anisotropy_2013, Steve_anisotropy_2013, Romano_anisotropy_2015, Gair_anisotropy_2015, Steve_bumpy_background, Chiara_anisotropy_2020, Nihan_anisotropy_forecast, nanograv_15yr_anisotropy}, thus relaxing assumption ii). One way of relaxing the assumption of Gaussianity is by modeling the background as a t-process, where the GWB power at each frequency follows a Student's t-distribution instead of a Gaussian distribution, thus allowing more flexibility (see Appendix D in \citealt{nanograv_15yr_gwb}). While such generalized search algorithms are available, the flagship results in most GWB searches continue to rely upon these assumptions (see e.g.~\citealt{nanograv_15yr_gwb, epta_dr2_gwb, ppta_dr3_gwb}). This is in part due to their simplicity, which makes them easier to compute. In addition, it is also widely expected that their assumptions are good enough to make them strong and robust methods.

The expectation that models employing these assumptions are capable of detecting GWBs, even when these assumptions are broken is based largely on \citet{Cornish:2013aba} and \citet{cornish_sampson_2016}. In \citet{Cornish:2013aba} it was shown that even a single SMBHB produces HD correlations in an array of isotropically distributed pulsars - showing that we can trade isotropy of the background for isotropy of the pulsar array. In \citet{cornish_sampson_2016} the authors studied both realistic GWBs based on SMBHB population models and also idealized Gaussian, isotropic GWBs with a power-law spectrum. They found that the significance with which one can detect these is similar, except for cases with an unrealistically low number of binaries. In this paper we carry out a similar analysis with a number of improvements:
\begin{enumerate}
    \item We use a more up-to-date SMBHB population model;
    \item Instead of pulsars with evenly sampled data, we use the actual observation times from the NANOGrav 15-year dataset \citep{nanograv_15yr_dataset};
    \item In addition to white noise, we include pulsar red noise as well, and we set the level of those based on the noise properties found in the NANOGrav 15-year dataset \citep{nanograv_15yr_detchar};
    \item Instead of simple frequency-domain analysis, we use the actual software used in \citet{nanograv_15yr_gwb}, that works in the time-domain, and takes into account covariances with the timing model.
\end{enumerate}

Our analysis can also be considered as an extension of the consistency checks carried out in \citet{nanograv_15yr_code_review}. There, these analysis pipelines were extensively stress tested to make sure they perform properly under the assumption that the data analyzed conforms to the models used. We extend the scope to see how these pipelines perform if their assumptions are not met perfectly, as is expected to be the case when analyzing real data. This is a special kind of model misspecification, where our signal model does not match the data. Understanding the effects of signal model misspecification, along with noise model misspecification \citep{goncharov2021} and developing consistency checks for our models \citep{Pat_post_Pred_check}, will be increasingly important as PTAs get more and more sensitive.

The rest of the paper is organized as follows. In Section \ref{sec:methods}, we describe our realistic population-based simulated datasets and give more details about the statistical methods employed on them. In Section \ref{sec:results}, we present our results, and in Section \ref{sec:conclusion} we provide a brief conclusion and discuss future work.

\section{\label{sec:methods}Methods}

\subsection{\label{sec:simulations}Realistic simulated datasets}
Our simulated datasets are based on the pulsars and their measured noise properties in the NANOGrav 15-year dataset \citep{nanograv_15yr_dataset}, following the simulation framework of \citet{astro4cast}. We use the \texttt{libstempo} software package, and we simulate datasets with the actual observing times from the real dataset, but to reduce the data volume we only keep one observation per epoch. We simulate data with the white and red noise properties fixed to the  maximum \textit{a posteriori} values inferred from the NANOGrav 15-year dataset \citep{nanograv_15yr_detchar}.

To simulate the contribution of a realistic GWB, we use an SMBHB population model implemented in the \texttt{holodeck} software package \citep{holodeck}, based on the Illustris cosmological hydrodynamical simulation \citep{Illustris}. This model applies a post processing step to account for small-scale physics not captured by Illustris, uses a phenomenological model to describe binary evolution (for details see \citealt{nanograv_15yr_astro}), and produces a simulated list of binaries in the Universe \citep{Luke_paper1, Luke_paper2, Luke_single_source}. To add the contribution of these binaries, we model the brightest 1000 binaries in each frequency bin individually and add the rest as a stochastic background. This was shown to produce equivalent results to a full simulation, but with drastically reduced runtimes \citep{Becsy:2022pnr}.
We produce hundreds of realizations of these simulations by drawing new binary populations to account for cosmic variance, and randomizing their extrinsic parameters (sky location, inclination, polarization angle, phases). We then analyze them with both Bayesian and frequentist statistical methods, which are described in the next sections.

\subsection{\label{sec:bayesian_methods}Bayesian methods}
To carry out a Bayesian statistical analysis of our datasets, we use the following likelihood function (for more details see, e.g.,~Eq.~(7.35) in \citealt{steve_book}):
\begin{equation}
    \log L = -\frac{1}{2} \left[ \delta t^T C^{-1} \delta t + \log \det (2 \pi C)   \right],
\end{equation}
where $\delta t$ are the timing residuals, and $C$ is the noise covariance matrix that includes the effects of both white and red noise and the linearized timing model. The matrix $C$ depends on parameters describing the noise. In our analysis we fix the white noise parameters to their true values. We model the spectrum of intrinsic pulsar red noise with a power-law model, so that its power spectral density (PSD) in the $i$th pulsar is:
\begin{equation}
    S_i(f) = \frac{A_i^2}{12 \pi^2} \left( \frac{f}{f_{\rm yr}} \right)^{-\gamma_i} \ f_{\rm yr}^{-3},
\end{equation}
where $A_i$ and $\gamma_i$ are the amplitude and the spectral slope of the RN in the $i$th pulsar, respectively, and $f_{\rm yr} = ({\rm yr})^{-1}$ is the frequency corresponding to a period of a year. We marginalize over $A_i$ and $\gamma_i$ in our analysis. Similarly, we model the timing residuals induced by the GWB as a red noise process with a power-law spectrum, such that its cross-power spectral density between the $i$th and $j$th pulsar is:
\begin{equation}
    S_{ij}(f) = \Gamma_{ij} \frac{A_{\rm GWB}^2}{12 \pi^2} \left( \frac{f}{f_{\rm yr}} \right)^{-\gamma_{\rm GWB}} \ f_{\rm yr}^{-3},
\end{equation}
where $\Gamma_{ij}$ is the overlap reduction function (ORF) between the two pulsars. For a GWB, the ORF is given by the HD curve \citep{HD}, which introduces correlations between pulsars depending on their angular separation on the sky. Another useful model is a so called common uncorrelated red noise (CURN), which assumes $\Gamma_{ij}=\delta_{ij}$, so no correlation between pulsars. This model is much more efficient to calculate, because $C$ is block-diagonal. In practice, the posteriors on $A_{\rm GWB}$ and $\gamma_{\rm GWB}$ are similar between the CURN and HD models. This allows one to sample the simpler CURN model and reweight the posterior samples to get fair draws from the HD posterior (for details see \citealt{sophie_resampling}). The mean of the weights also serves as a measurement of the BF between the HD and CURN models, which is the primary quantity one needs to measure to claim a detection of the GWB. The variance of the weights can be used to approximate the number of effectively independent samples after reweighting. The ratio of this to the total number of samples defines the reweighting efficiency, which needs to be sufficiently high to get reliable results. Over the 773 realizations we analyzed, only 17 showed a reweighting efficiency of less than 5\%, most of which resulted only in a modest error of the measured BFs, despite the low efficiency. Thus the use of the reweighting technique should not have a significant effect on our results.

\subsection{\label{sec:frequentist_methods}Frequentist methods}
While GWB searches usually focus on Bayesian methods, frequentist methods can also be useful, particularly given their significantly faster computational speed. The most common frequentist method used in searches for a GWB is the so called optimal statistic (OS, \citealt{OS}), which is an unbiased estimator of the GWB amplitude. It is constructed by maximizing the likelihood of the measured cross-correlation values under the assumption of an ORF and a (typically power-law) PSD model. This results in the estimated GWB amplitude ($\hat{A}$) and the standard deviation of that estimate ($\sigma_A$). The ratio of these gives the signal-to-noise ratio, S/N, which can be used as a detection statistic for the presence of a GWB.

One disadvantage of the OS is that one needs to know the PSD describing the GWB and the intrinsic red noise of each pulsar, which is not known \textit{a priori} in a real dataset. One can measure the PSD using Bayesian methods, and use a point estimate (usually the maximum \textit{a posteriori}) to calculate the OS. However, this does not take into account the uncertainty of the measured spectrum. To do so, \citet{nmos} developed the so called noise-marginalized OS, which calculates the OS for random draws of the spectrum from a Bayesian posterior, thus producing a distribution of OS values. Subsequently, \citet{Michele_post_pred_check} have also proposed a related method for interpreting the OS in the context of posterior predictive checking.

The OS can also be used to calculate the S/N of processes described by different ORFs. Two commonly calculated ORFs are the monopole and dipole, which could arise e.g.~from systematic clock errors and ephemeris errors, respectively. One limitation of the standard OS, is that since these ORFs are usually not orthogonal, processes with less preferred ORFs can still produce high S/N values. To alleviate this problem, \citet{mcos} developed the so called multi-component OS, which can allow multiple processes with different ORFs to fit the data simultaneously. As a result, different processes can be better decoupled. In this study we calculate S/N values with the noise-marginalized multi-component OS method to assess the frequentist significance of the GWB, along with monopole or dipole correlated processes.

\section{\label{sec:results}Results}
\begin{figure}[htbp!]
    \centering
    \includegraphics[width = 1\columnwidth]{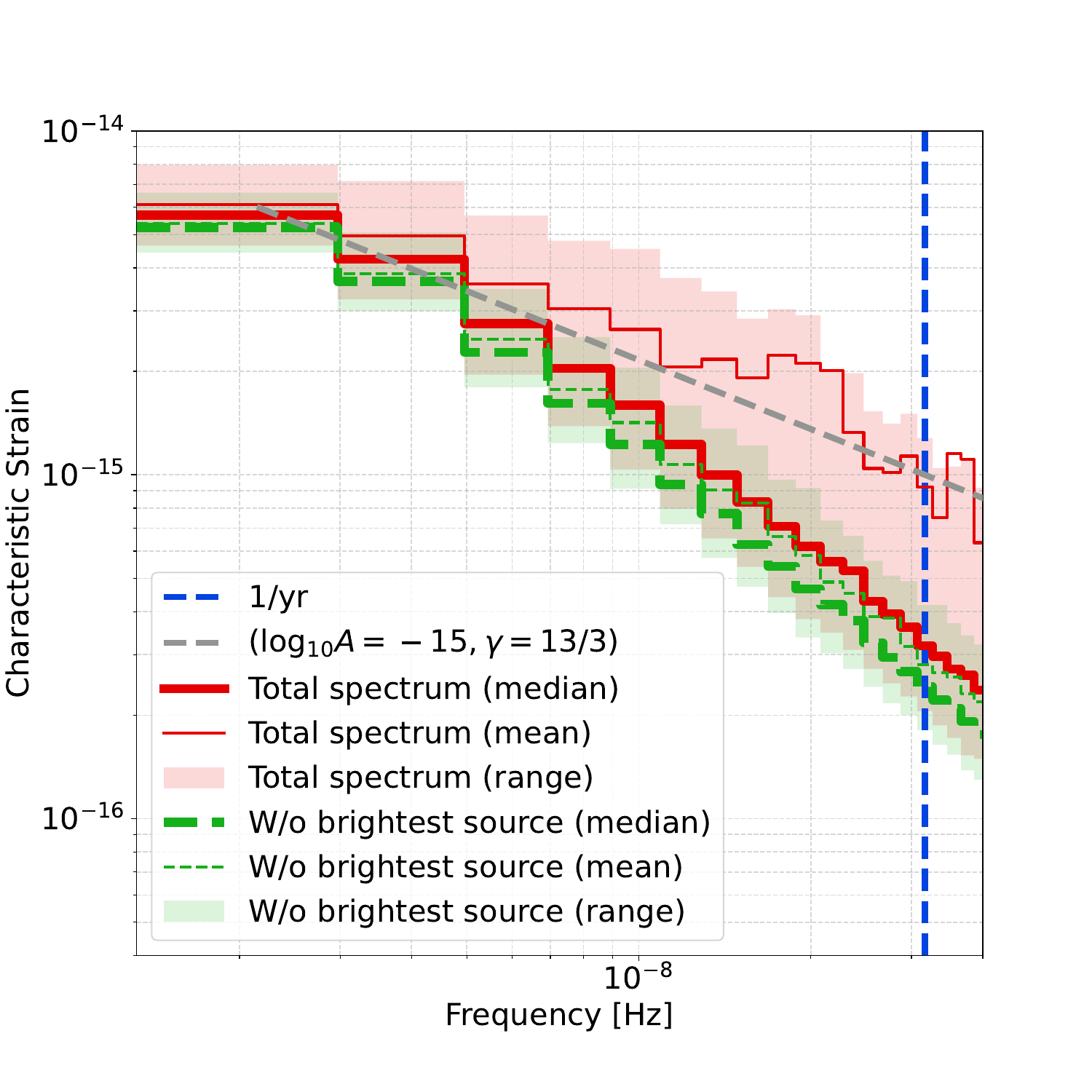}
    \caption{Mean and median GWB spectra over 773 realizations. Also shown is the spectrum after removing the brightest binary's contribution in each frequency bin. The shaded regions show the range of spectra between the 5th and 95th percentile. We also show a power-law spectrum with a canonical 13/3 slope as a reference.}
    \label{fig:spectrum}
\end{figure}

To produce robust results, we analyzed 773 different realizations of our simulated 15-yr dataset. The median (thick lines) and mean (thin lines) GWB spectra over these realizations is shown in Figure \ref{fig:spectrum}, both for all the binaries (red solid lines) and for all except the brightest binaries in each bin (green dashed lines). The shaded region represents amplitudes between the 5th and 95th percentile in each bin. We also show a spectrum with $\gamma=13/3$ as a reference. Note that this has a slope between the median and mean lines for all sources, which give $\gamma=4.6$ and $\gamma=4.2$ in a linear fit to the first five frequency bins, respectively. We can also see that the median spectrum is similar to the mean spectrum once the brightest binary is removed. However, for the total spectrum, the mean spectrum is shallower and follows the expected 13/3 spectral slope better. This is due to the fact that the median is insensitive to outliers, and thus we lose the information about the bright sources.

\subsection{\label{ssec:gwb}Stochastic background}

\begin{figure}[htbp!]
    \centering
    \includegraphics[width = 1\columnwidth]{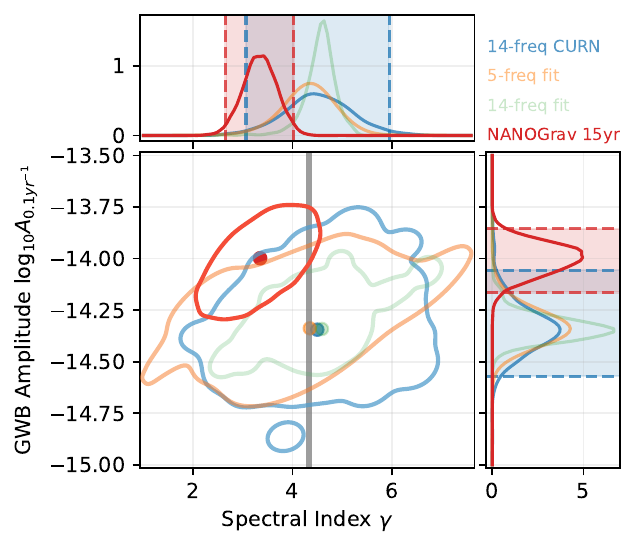}
    \caption{Distribution of recovered median $\log_{10} A_{0.1 \mathrm{yr}^{-1}}$ and $\gamma_{\rm GWB}$ values from 14-frequency CURN analysis of 773 realizations (blue). Also shown are the values we get from a simple least-squares fit to the first 5 (orange) and first 14 (green) frequency bins of the simulated spectra, and the parameters reported in \citet{nanograv_15yr_gwb} based on the NANOGrav 15-year dataset (red). Contours represent 99.7\% levels (3$\sigma$), and dots show the medians. 1-dimensional marginal distributions are also shown, with 95.4\% intervals (2$\sigma$) highlighted for some of the distributions. While the NANOGrav results show a $\gamma_{\rm GWB}$ value lower than the $13/3$ value expected from circular GW-driven GWB (gray), it is well within the spread of $\gamma_{\rm GWB}$ values found in our simulated datasets.}
    \label{fig:A_gamma_scatter}
\end{figure}

We carry out a Bayesian search for a 14-frequency CURN process in each realization, resulting in posterior distributions of the power-law parameters. Based on those, we calculate the median $A_{\rm GWB}$ and $\gamma_{\rm GWB}$ values for each realization. The distribution of these are shown on Figure \ref{fig:A_gamma_scatter} (blue), along with $A_{\rm GWB}$ and $\gamma_{\rm GWB}$ values estimated by a simple least squares fit to the theoretical spectrum using either the first 5 (orange) or the first 14 (green) frequency bins. Interestingly, fitting to the first 5 frequencies gives very similar values to the actual Bayesian CURN analysis, even though that uses the first 14 frequencies. This is due to the fact that the Bayesian analysis is dominated by the first few frequencies, where the signal is the strongest. Conversely, a 14-frequency fit to the spectra provides unrealistically accurate measurements, as it does not take into account the effect of red and white noise, the latter of which dominates the GWB signal at higher frequencies. Note that all of these distributions center roughly around the theoretically expected $\gamma_{\rm GWB}=13/3$ value, and the amplitude to which this population was calibrated. However, they show a significant spread indicating that the lower $\gamma_{\rm GWB}$ value and higher amplitude found in the NANOGrav 15-year GWB search (posterior distribution and median values shown by red contour and dot on Figure \ref{fig:A_gamma_scatter}) is compatible with the binary population in our simulations. Note that this is consistent with the findings of \citet{nanograv_15yr_gwb}, where they found that the NANOGrav results are compatible with an SMBHB interpretation based on a comparison between the recovered $A_{\rm GWB}$ and $\gamma_{\rm GWB}$ values and fits to SMBHB population models.\footnote{Note that \citet{nanograv_15yr_astro} carried out much more rigorous analyses, which also showed that the GWB is consistent with an SMBHB interpretation.} As we can see, Bayesian spectral analysis of population-based simulated datasets yield similar conclusions. This is not surprising given that we see a simple fit to the spectrum gives very similar results to a full analysis. 

\begin{figure}[bp!]
    \centering
    \includegraphics[width = 1\columnwidth]{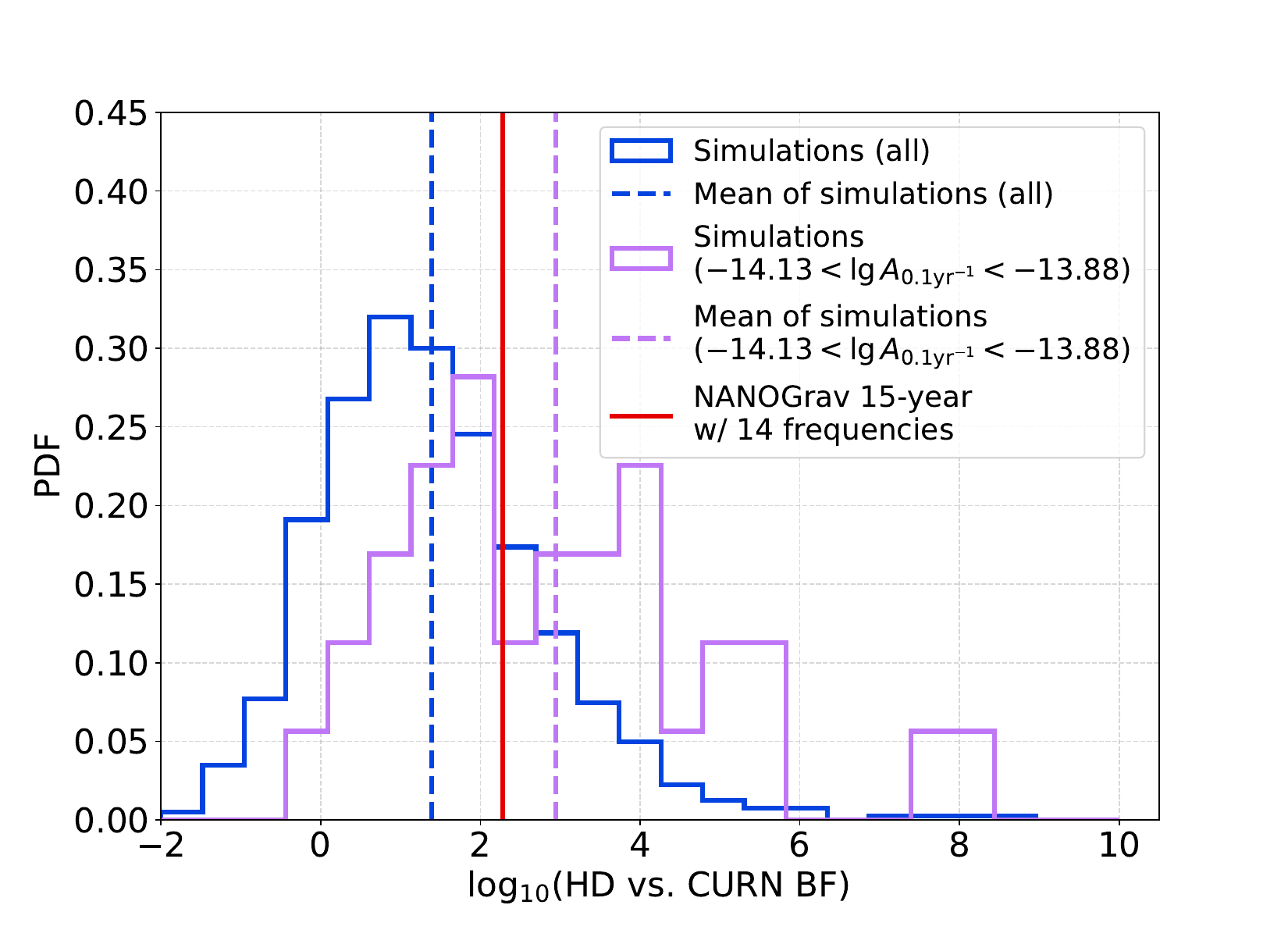}
    \caption{Distribution of HD vs.~CURN BFs for all 773 realizations (blue) and for 14 realizations with a recovered amplitude consistent with the amplitude recovered in the NANOGrav 15-year analysis at the 90\% confidence level (purple). Also shown are the means of these distributions (dashed vertical lines), and the BF reported in \citet{nanograv_15yr_gwb} based on the NANOGrav 15-year dataset (red). While this particular astrophysical model predicts amplitudes and BFs that tend to be lower than the one found in the NANOGrav 15-year dataset, if we select a subset of these that produce a consistent amplitude, they also show good consistency in terms of BF.}
    \label{fig:BF_histogram}
\end{figure}

\begin{figure}[htbp!]
    \centering
    \includegraphics[width = 1\columnwidth]{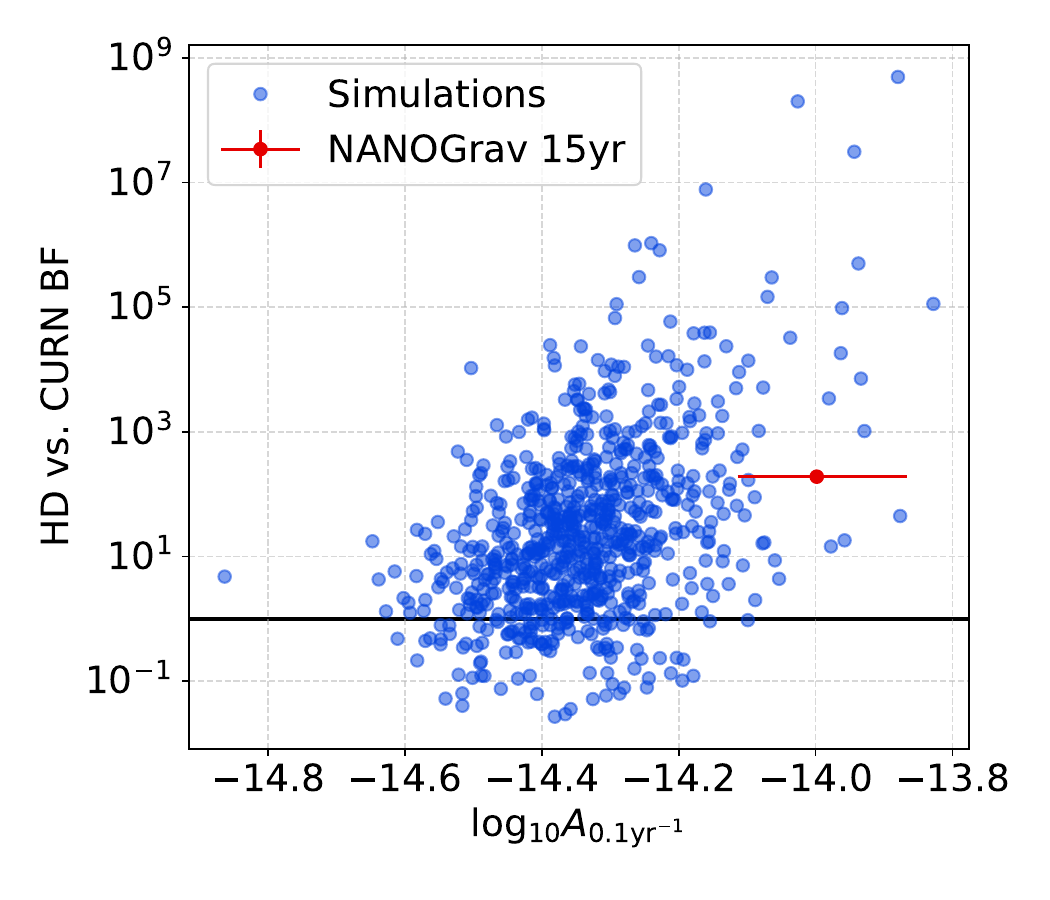}
    \caption{HD vs.~CURN BFs as a function of the recovered GWB amplitude referenced at $f=0.1 {\rm yr}^{-1}$ (blue dots). While there is correlation between the amplitude and the BF, the BFs show significant variance even at a given amplitude. Also shown is the BF and amplitude reported in \citet{nanograv_15yr_gwb} based on the NANOGrav 15-year dataset (red).}
    \label{fig:BF_vs_amplitude}
\end{figure}

For each realization, we also produce HD posteriors using the reweighting method \citep{sophie_resampling}. We calculate the BF between the HD and CURN models which are shown as the blue histogram in Figure \ref{fig:BF_histogram}. Note the large variance of the BF, even though we fixed the astrophysical parameters of the simulation (more than eight orders of magnitude between the lowest and highest BF). To understand the source of this variance, it is worth looking at the recovered BFs as a function of the amplitude of the GWB. Figure \ref{fig:BF_vs_amplitude} shows the BFs as a function of $\log_{10} A_{0.1 \mathrm{yr}^{-1}}$, i.e.~the amplitude of the GWB referenced at a frequency of $0.1 \mathrm{yr}^{-1}$ instead of the usual $1 \mathrm{yr}^{-1}$. We use this reference point, since there the amplitude is less covariant with $\gamma_{\rm GWB}$. We can see that while the BF is correlated with the amplitude, even at a given fixed amplitude we see significant scatter in the BF. Thus even for universes described by the same astrophysics and a GWB with the same amplitude, the recovered BF shows a significant variance. This is consistent with previous findings on both astrophysical and simple power-law backgrounds \citep{cornish_sampson_2016, sophie_resampling}. Thus this is a generic property of any stochastic background, not specifically of astrophysical backgrounds.

\begin{figure*}[htbp!]
    \centering
    \includegraphics[width = 2\columnwidth]{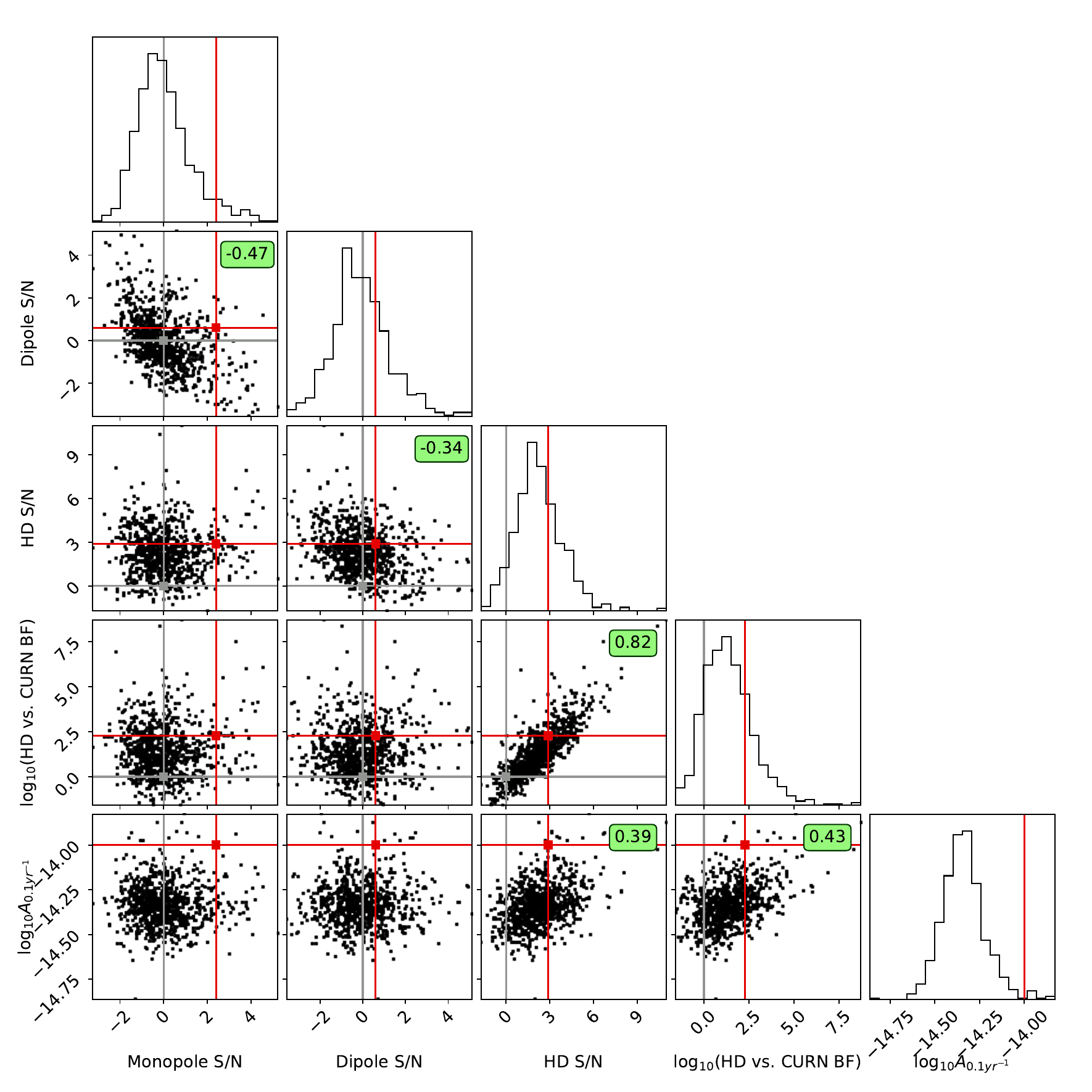}
    \caption{Distribution of mean OS S/Ns, HD vs.~CURN BFs, and recovered GWB amplitude referenced at $f=0.1 {\rm yr}^{-1}$. Grey lines represent zero S/Ns and unit BFs, and red lines represent the values reported in \citet{nanograv_15yr_gwb} based on the NANOGrav 15-year dataset. We also indicate the Pearson correlation coefficient for every pair where the associated p-value is below $10^{-3}$.}
    \label{fig:BF_OS_Amp_corner}
\end{figure*}

Figure \ref{fig:BF_histogram} also shows the histogram of BFs for realizations that show an amplitude consistent at the 90\% level with that reported in the NANOGrav 15-year GWB search \citep{nanograv_15yr_gwb}, along with the actual measured BF in that search. We can see that the simulations show about four orders of magnitude scatter even in this narrow amplitude range, but they predict BFs that are roughly consistent with the real data. Note that this is not a rigorous consistency check of the 15-year results with the astrophysical predictions, since our astrophysical models tend to produce a lower amplitude than the one found in the 15-year dataset. Thus, the mean of our simulations produces a lower amplitude and lower BF than found in the 15-year data. However, once we filter out only those realizations that produce an amplitude close to the one found in the real data, we find that the BFs are consistent with the one measured on real data. This stands to show that our simplified analysis methods are capable of detecting the GWB with a significance reported in \citep{nanograv_15yr_gwb} even if the GWB is coming from a finite number of binaries.

We also calculate the noise marginalized multi-component optimal statistic for all realizations. Figure \ref{fig:BF_OS_Amp_corner} shows the distribution of the mean monopole, dipole, and HD S/Ns, along with the HD vs.~CURN BFs and $\log_{10} A_{0.1 \mathrm{yr}^{-1}}$. We also indicate the values found in the NANOGrav 15-year dataset on Figure \ref{fig:BF_OS_Amp_corner} (red markers). Note that these are all roughly consistent with our simulations. Black lines represent zero S/Ns and BFs. Note that as expected both monopole and dipole S/N values are centered around zero, while the HD S/N distribution is offset towards positive values.

\begin{figure*}[htbp!]
    \centering
    \includegraphics[width = 1.5\columnwidth]{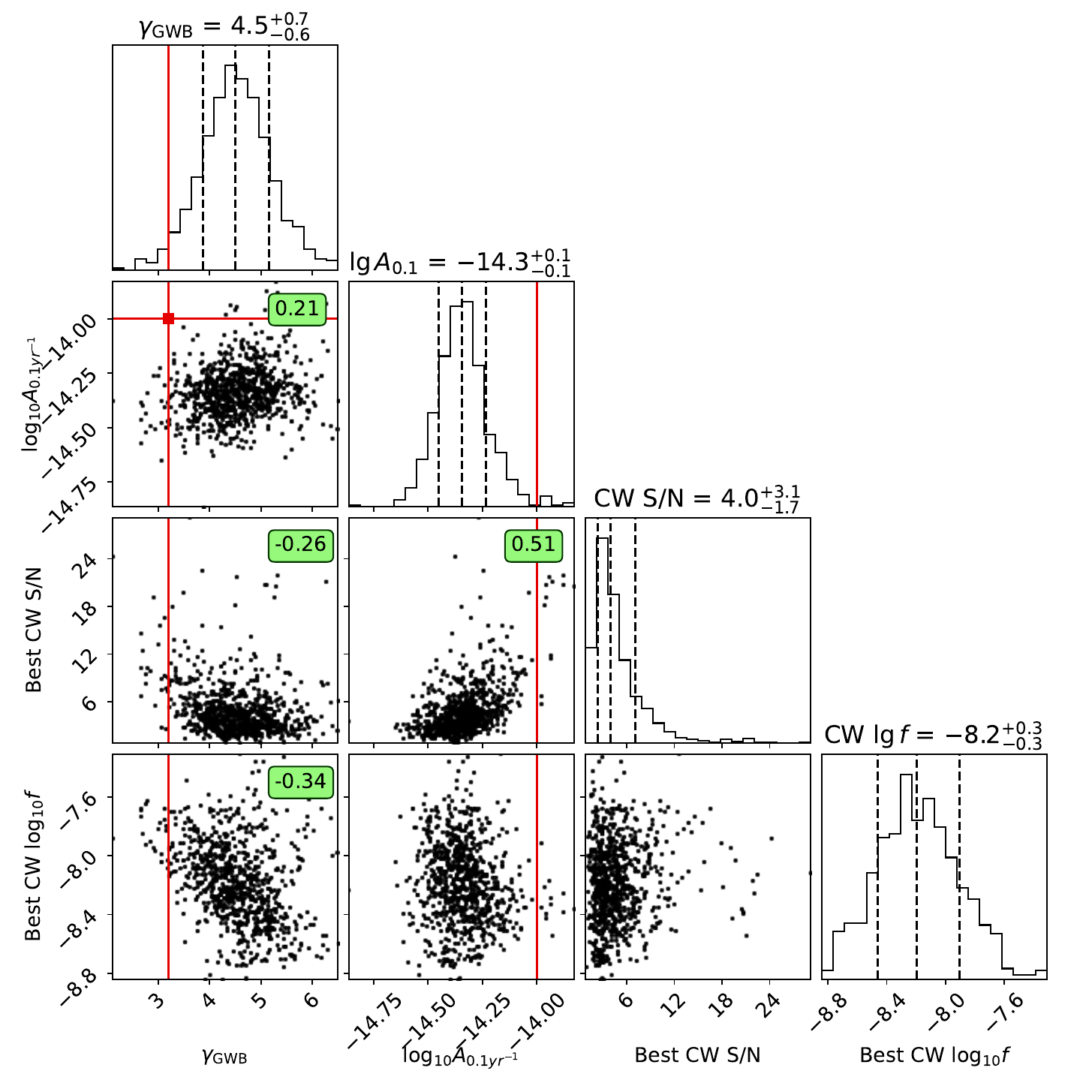}
    \caption{Distribution of GWB spectrum parameters ($\gamma_{\rm GWB}$, $\log_{10} A_{0.1 \mathrm{yr}^{-1}}$) based on CURN analysis, and S/N and $\log_{10} f$ of the highest-S/N CW source in each realization. Red lines represent the values reported in \citet{nanograv_15yr_gwb} based on the NANOGrav 15-year dataset. We also indicate the Pearson correlation coefficient for every pair where the associated p-value is below $10^{-3}$.}
    \label{fig:gwb_cw_corner}
\end{figure*}

We can see some interesting correlations between values shown on Figure \ref{fig:BF_OS_Amp_corner}. To quantify these, we calculate the Pearson correlation coefficient ($r$) for each parameter pair and show them on the figure where the correlation is significant (p-value $< 10^{-3}$). The strongest correlation of $r=0.82$ can be observed between HD S/N and HD vs.~CURN BF. This is reassuring, since these are both meant to quantify the significance of the presence of HD cross-correlations in the data. In fact, both \citet{astro4cast} and \citet{Michele_post_pred_check} gives an approximate mapping between these two quantities. Both the HD S/N and HD vs.~CURN BF values are also correlated with  $\log_{10} A_{0.1 \mathrm{yr}^{-1}}$, with $r=0.39$ and $r=0.43$, respectively. These relatively low correlation values are in line with our finding above, that a large variation remains in detection statistics even at fixed amplitude. The fact that the amplitude is more correlated with the BF than with the HD S/N can potentially be explained by the the fact that the Bayesian analysis uses uses both auto-correlation and cross-correlation information, while the OS relies solely on cross-correlations. Additionally, both monopole and HD S/N is anticorrelated with the dipole S/N. However, there seems to be no correlation between the HD and monopole S/N.

\subsection{\label{ssec:cw}Individual binaries}

We investigate detection prospects of individual binaries and how they relate to background properties by following \citet{Becsy:2022pnr}: we calculate the expected S/N of binaries defined as $\mathrm{S/N}=\sqrt{(s|s)}$, where $s$ is the simulated CW signal, i.e.~we calculate the inner product of the waveform with itself. We do so for each binary with randomly assigned extrinsic parameters (sky location, inclination angle, polarization angle, phases), and we find the binary with the highest S/N in each realization. Figure \ref{fig:gwb_cw_corner} shows the S/N and GW frequency of that best binary in each realization, along with the $\log_{10} A_{0.1 \mathrm{yr}^{-1}}$ and $\gamma_{\rm GWB}$ values we find in our CURN analysis. Similarly to \citet{Becsy:2022pnr} we find that the the highest-S/N sources tend to be found at moderate frequencies around 2-10 nHz. We find a higher mean S/N, which is not surprising given the inclusion of additional pulsars and updated noise models of NANOGrav 15-year pulsars.

We calculate the Pearson correlation coefficient ($r$) for each parameter pair and show it on Figure \ref{fig:gwb_cw_corner} where the correlation is significant (p-value $< 10^{-3}$). The most significant correlation is shown between the CW S/N and $\log_{10} A_{0.1 \mathrm{yr}^{-1}}$, with $r=0.51$. This can be attributed to the fact that the CURN analysis does not include a CW model, so if a CW is present, the recovered amplitude is expected to be biased high. We also find significant anti-correlation between the CW S/N and $\gamma_{\rm GWB}$ ($r=-0.26$). This is probably due to the fact that $\gamma_{\rm GWB}$ can be biased low by a significant individual binary at moderate frequencies as we will see below. Finally, the only parameter that correlates with the frequency of the loudest binary is $\gamma_{\rm GWB}$, showing a negative correlation with $r=-0.34$. This is understandable by the fact that the higher the binary's frequency is, the more it can bias the GWB slope low.

To further investigate the covariance between individual binaries and the GWB, we analyzed a single realization with the model of an individual binary and CURN (CURN+CW). For this analysis we used \texttt{QuickCW} \citep{QuickCW, QuickCW_code}, a software package that builds on the \texttt{enterprise} \citep{enterprise} and \texttt{enterprise\textunderscore extensions} \citep{ee} libraries, but uses a custom likelihood calculation and a Markov chain Monte Carlo (MCMC) sampler tailored to the search for individual binaries. We selected this realization randomly under the constraints that it has a CW S/N$>$5 and a monopole S/N$>$1.5. The former ensured that there is a detectable individual binary in the data, while the latter was motivated by the subdominant monopolar signature found around 4 nHz in the NANOGrav 15-year dataset \citep{nanograv_15yr_gwb}, which was also identified by the individual binary search carried out using that dataset \citep{nanograv_15yr_cw}. The selected realization has a monopole S/N of 1.8, a dipole S/N of $-1.33$, an HD S/N of 2.29, an HD vs CURN BF of 77, a $\log_{10} A_{0.1 \mathrm{yr}^{-1}}= -14.4$, a CW SNR of 6.2, and a CW frequency of 9 nHz.

\begin{figure}[htbp!]
    \centering
    \includegraphics[width = 1\columnwidth]{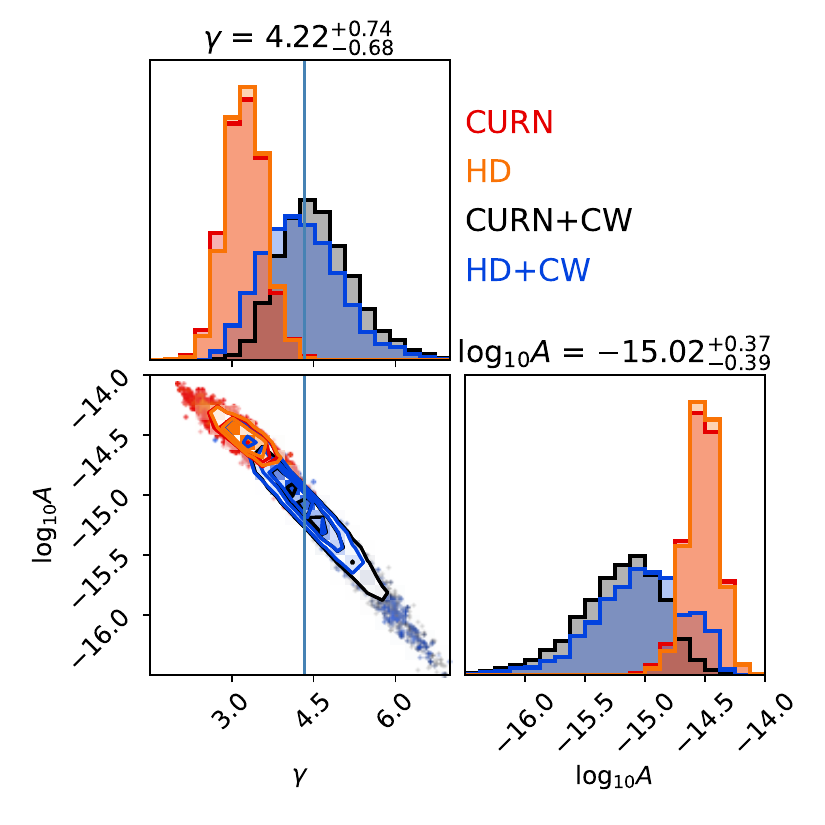}
    \caption{Distribution of GWB parameters under the CURN (red), HD (orange), CURN+CW (black), and HD+CW (blue) models for a particular realization with a clearly detectable binary ($\mathrm{S/N}=6.2$). Not modeling the CW signal biases the spectral slope low, while once we include it we recover the expected $\gamma=13/3$ value. Note that in this particular example, the CURN and HD models show very similar posteriors, while the spectral recovery is slightly different between the CURN+CW and HD+CW models.}
\label{fig:gwb_w_and_wo_cw_corner}
\end{figure}

Analyzing this dataset with the CURN+CW model we found that the CW signal was clearly detected, with a CURN+CW vs CURN BF of $\sim$25. Moreover, the parameter recovery of the CURN process was significantly affected by the inclusion of the CW model. Figure \ref{fig:gwb_w_and_wo_cw_corner} shows the distribution of $\gamma_{\rm GWB}$ and $\log_{10} A$ under the CURN (red), HD (orange), CURN+CW (black), and HD+CW (blue) models. Posteriors with models including HD were produced by reweighting \citep{sophie_resampling}. We can see that in the case of the CURN-only and HD-only models, the unmodeled binary biases the $\gamma_{\rm GWB}$ recovery low compared to the canonical 13/3 value, while the CURN+CW and HD+CW models produce $\gamma_{\rm GWB}$ posteriors centered on values close to 13/3. This is understandable, as the CURN model can partially model the CW signal by introducing more power at higher frequencies where the CW is present. This is also consistent with the correlation between $\gamma_{\rm GWB}$ and CW properties we found on Figure \ref{fig:gwb_cw_corner}. While we cannot draw definitive conclusions from a single realization, this result suggests that modeling individual binaries along with the GWB may be crucial for unbiased spectral characterization of the GWB. We leave a detailed investigation of this problem to a future study.

It is important to note that while the NANOGrav 15-year analysis recovers a $\gamma_{\rm GWB}$ value lower than 13/3, that does not significantly change with the inclusion of an individual binary model \citep{nanograv_15yr_cw}, and the $\gamma_{\rm GWB}$ recovery is instead more affected by the particular noise models employed \citep{nanograv_15yr_gwb}. This is unlike the EPTA DR2 analysis, where the inclusion of an individual binary completely changes the GWB recovery \citep{epta_dr2_cw}. This highlights that both a misspecified signal model and noise model can have adverse effects on parameter estimation.

Note that in this particular realization, the recovered spectra under the CURN and HD models are practically identical (red and orange on Figure \ref{fig:gwb_w_and_wo_cw_corner}). This is not particularly surprising, as this forms the basis of the reweighting technique \citep{sophie_resampling}. In the presence of an individual binary, the inclusion of correlations does have a small effect on the recovered spectrum (compare black and blue on Figure \ref{fig:gwb_w_and_wo_cw_corner}), but the difference is small enough that the reweighting technique can be effective.

\begin{figure*}[htbp]
    \centering
    \includegraphics[width = 2\columnwidth]{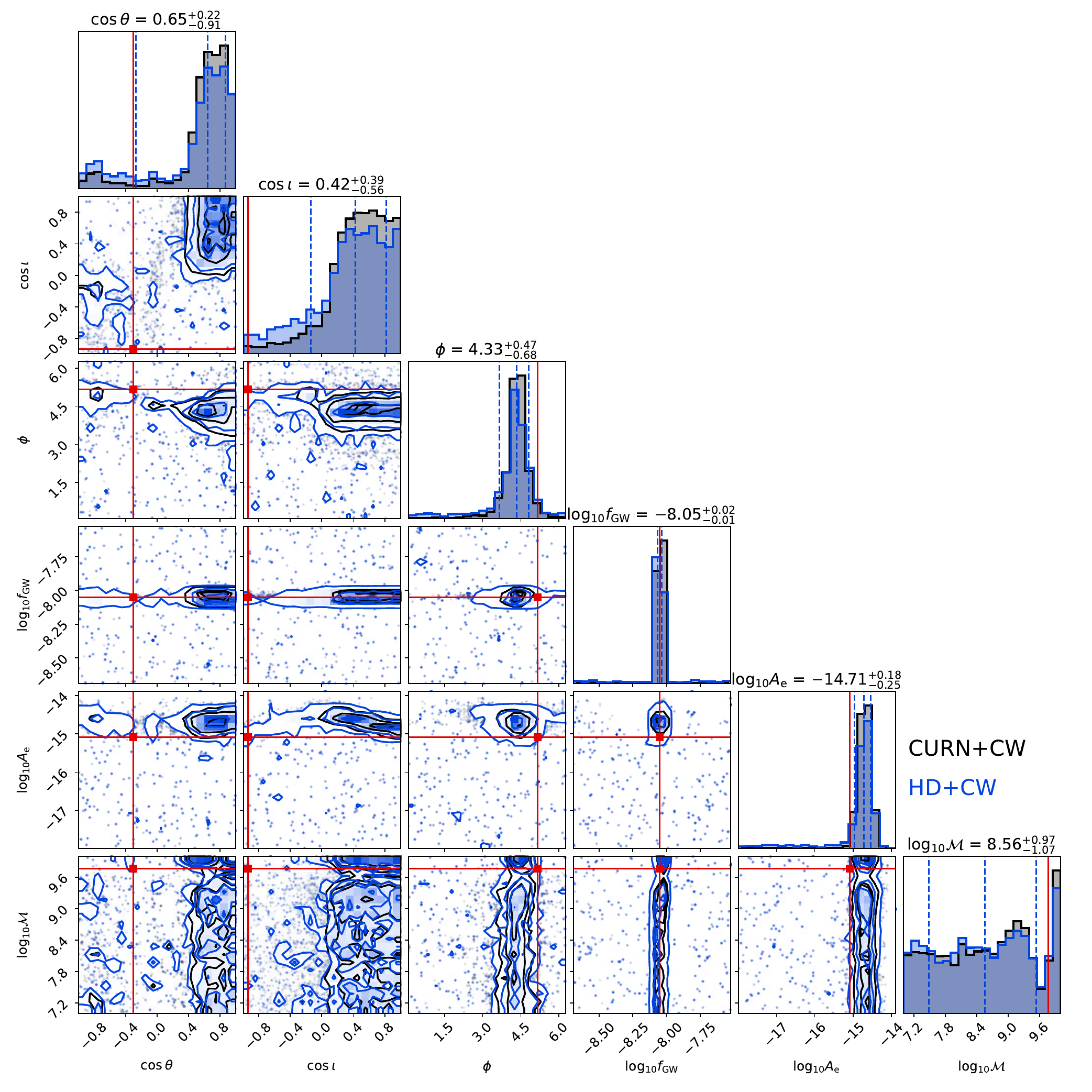}
    \caption{Distribution of CW parameters for a particular realization from a CURN+CW analysis (black) and those reweighted to HD+CW (blue). Red lines represent the true parameter values.}
    \label{fig:cw_corner_reweight}
\end{figure*}

In addition, the significance of the individual binary does not change significantly after reweighting. We show the distribution of some individual binary model parameters on Figure \ref{fig:cw_corner_reweight}. Black shows posteriors from the CURN+CW run, while blue shows those reweighted to HD+CW. Red indicates the true parameters of the signal. In both cases, the binary is clearly detected. The reweighted posterior distributions are similar to the original ones, but are less constrained. We can also see that the recovery of some parameters show non-negligible bias. This run is well-converged, and \texttt{QuickCW} has been shown to produce unbiased parameter estimates \citep{nanograv_15yr_cw} on CURN+CW simulations. Thus we suspect this bias is due to model mis-specification, e.g.~because the background is modeled with a power-law spectrum, or because we assume there is only one significant binary present. In particular, the unmodeled contributions of subdominant binaries at similar frequencies could potentially result in strong biases in some parameters as the sampler tries to adjust the model to fit contributions from more than one binary. In this particular realizations there are three binaries with frequencies within $1/(2T_{\rm obs})\simeq2$ nHz, and amplitudes within a factor of three of the loudest binary. This effect is being investigated in a follow-up systematic injection study, where we analyze a large number of realizations under the CURN+CW model. If this bias is indeed the result of the single-binary assumption, an analysis modeling multiple binaries simultaneously could alleviate the problem. An  efficient implementation of such an analysis is under development based on an existing multi-binary pipeline (\texttt{BayesHopper}, \citealt{BayesHopper}) and techniques from the efficient single-binary \texttt{QuickCW} algorithm \citep{QuickCW}.

\section{\label{sec:conclusion}Conclusion and future work}
We used state-of-the art statistical data analysis techniques to search for a stochastic GW background in realistic simulated datasets. We employed both Bayesian and frequentist techniques to estimate the parameters of the background and assess its significance. Our simulated datasets were produced based on an astrophysical population of supermassive black hole binaries, and as such the resulting stochastic backgrounds were anisotropic, non-Gaussian, and had non-power-law spectra. We found that standard analysis techniques were able to detect the characteristic Hellings-Downs correlations in these backgrounds, despite the fact that they assume isotropic, Gaussian, power-law backgrounds.

These analyses were also able to correctly characterize the spectrum of these backgrounds most of the time. However, we found the presence of a strong individual binary can bias the recovered amplitude high, and the recovered spectral slope low. We calculated the signal-to-noise ratio of the loudest individual binary in each realization and examined how that relates to the background properties.

The fact that standard statistical analysis techniques used by the PTA community can indeed detect the signals from a realistic SMBHB population is reassuring. On the other hand, the possibility that individual binaries can bias spectral recovery is cause for concern, and suggests that joint analyses of individual binaries and the background might become necessary to avoid biased inference. Future studies investigating the interplay between these two source types will be crucial in making sure that any astrophysical interpretation is robust. In addition, it will be important to assess the importance of modeling multiple binaries simultaneously. We also plan to investigate the expected level of anisotropy in the stochastic background based on these realistic simulations using methods described e.g.~in \citet{nanograv_15yr_anisotropy} and \citet{Emiko_cw_paper}. While the results presented in this paper are based on a fixed astrophysical model, investigating the effect of astrophysical parameters on the resulting background and individual binary prospects is also being investigated \citep{Emiko_cw_paper}.

Note that a similar independent analysis was reported concurrently in \citet{valtolina2023}, with results consistent with ours, based on simulated datasets resembling the latest dataset of the European Pulsar Timing Array. This highlights that our results are robust against the choice of the specific dataset we base our simulations on. One difference between the results is that \citet{valtolina2023} finds that their spectral index recovery is biased high, while we see an unbiased recovery of the spectral index on average (see Figure \ref{fig:A_gamma_scatter}). This is potentially due to the fact that the dataset we consider is significantly longer, which allows access to lower frequencies, where the spectrum is less affected by finite-number effects.

\section{Acknowledgements}
\label{sec:acks}
We appreciate the support of the NSF Physics Frontiers Center Award PFC-1430284 and the NSF Physics Frontiers Center Award PFC-2020265.
L.B. acknowledges support from the National Science Foundation under award AST-1909933 and from the Research Corporation for Science Advancement under Cottrell Scholar Award No. 27553.
P.R.B. is supported by the Science and Technology Facilities Council, grant number ST/W000946/1.
S.B. gratefully acknowledges the support of a Sloan Fellowship, and the support of NSF under award \#1815664.
M.C. and S.R.T. acknowledge support from NSF AST-2007993.
M.C. and N.S.P. were supported by the Vanderbilt Initiative in Data Intensive Astrophysics (VIDA) Fellowship.
K.Ch., A.D.J., and M.V. acknowledge support from the Caltech and Jet Propulsion Laboratory President's and Director's Research and Development Fund.
K.Ch. and A.D.J. acknowledge support from the Sloan Foundation.
Support for this work was provided by the NSF through the Grote Reber Fellowship Program administered by Associated Universities, Inc./National Radio Astronomy Observatory.
Support for H.T.C. is provided by NASA through the NASA Hubble Fellowship Program grant \#HST-HF2-51453.001 awarded by the Space Telescope Science Institute, which is operated by the Association of Universities for Research in Astronomy, Inc., for NASA, under contract NAS5-26555.
Pulsar research at UBC is supported by an NSERC Discovery Grant and by CIFAR.
K.Cr. is supported by a UBC Four Year Fellowship (6456).
M.E.D. acknowledges support from the Naval Research Laboratory by NASA under contract S-15633Y.
T.D. and M.T.L. are supported by an NSF Astronomy and Astrophysics Grant (AAG) award number 2009468.
E.C.F. is supported by NASA under award number 80GSFC21M0002.
G.E.F., S.C.S., and S.J.V. are supported by NSF award PHY-2011772.
The Flatiron Institute is supported by the Simons Foundation.
S.H. is supported by the National Science Foundation Graduate Research Fellowship under Grant No. DGE-1745301.
The work of N.La. and X.S. is partly supported by the George and Hannah Bolinger Memorial Fund in the College of Science at Oregon State University.
N.La. acknowledges the support from Larry W. Martin and Joyce B. O'Neill Endowed Fellowship in the College of Science at Oregon State University.
Part of this research was carried out at the Jet Propulsion Laboratory, California Institute of Technology, under a contract with the National Aeronautics and Space Administration (80NM0018D0004).
D.R.L. and M.A.M. are supported by NSF \#1458952.
M.A.M. is supported by NSF \#2009425.
C.M.F.M. was supported in part by the National Science Foundation under Grants No. NSF PHY-1748958 and AST-2106552.
A.Mi. is supported by the Deutsche Forschungsgemeinschaft under Germany's Excellence Strategy - EXC 2121 Quantum Universe - 390833306.
The Dunlap Institute is funded by an endowment established by the David Dunlap family and the University of Toronto.
K.D.O. was supported in part by NSF Grant No. 2207267.
T.T.P. acknowledges support from the Extragalactic Astrophysics Research Group at E\"{o}tv\"{o}s Lor\'{a}nd University, funded by the E\"{o}tv\"{o}s Lor\'{a}nd Research Network (ELKH), which was used during the development of this research.
S.M.R. and I.H.S. are CIFAR Fellows.
Portions of this work performed at NRL were supported by ONR 6.1 basic research funding.
J.D.R. also acknowledges support from start-up funds from Texas Tech University.
J.S. is supported by an NSF Astronomy and Astrophysics Postdoctoral Fellowship under award AST-2202388, and acknowledges previous support by the NSF under award 1847938.
S.R.T. acknowledges support from an NSF CAREER award \#2146016.
C.U. acknowledges support from BGU (Kreitman fellowship), and the Council for Higher Education and Israel Academy of Sciences and Humanities (Excellence fellowship).
C.A.W. acknowledges support from CIERA, the Adler Planetarium, and the Brinson Foundation through a CIERA-Adler postdoctoral fellowship.
O.Y. is supported by the National Science Foundation Graduate Research Fellowship under Grant No. DGE-2139292.


\software{\texttt{enterprise} \citep{enterprise},
          \texttt{enterprise\textunderscore extensions} \citep{ee},
          \texttt{QuickCW} \citep{QuickCW_code},
          \texttt{corner} \citep{corner},
          \texttt{libstempo} \citep{libstempo},
          \texttt{tempo} \citep{tempo},
          \texttt{tempo2} \citep{tempo2},
          \texttt{PINT} \citep{pint},
          \texttt{matplotlib} \citep{matplotlib},
          \texttt{astropy} \citep{astropy, astropy:2013}
          }


\bibliography{realistic_background_detection}{}

\begin{thebibliography}{}
\expandafter\ifx\csname natexlab\endcsname\relax\def\natexlab#1{#1}\fi
\providecommand{\url}[1]{\href{#1}{#1}}
\providecommand{\dodoi}[1]{doi:~\href{http://doi.org/#1}{\nolinkurl{#1}}}
\providecommand{\doeprint}[1]{\href{http://ascl.net/#1}{\nolinkurl{http://ascl.net/#1}}}
\providecommand{\doarXiv}[1]{\href{https://arxiv.org/abs/#1}{\nolinkurl{https://arxiv.org/abs/#1}}}

\bibitem[{{Afzal} {et~al.}(2023){Afzal}, {Agazie}, {Anumarlapudi}, {Archibald},
  {Arzoumanian}, {Baker}, {B{\'e}csy}, {Blanco-Pillado}, {Blecha}, {Boddy},
  {Brazier}, {Brook}, {Burke-Spolaor}, {Burnette}, {Case}, {Charisi},
  {Chatterjee}, {Chatziioannou}, {Cheeseboro}, {Chen}, {Cohen}, {Cordes},
  {Cornish}, {Crawford}, {Cromartie}, {Crowter}, {Cutler}, {Decesar}, {Degan},
  {Demorest}, {Deng}, {Dolch}, {Drachler}, {von Eckardstein}, {Ferrara},
  {Fiore}, {Fonseca}, {Freedman}, {Garver-Daniels}, {Gentile}, {Gersbach},
  {Glaser}, {Good}, {Guertin}, {G{\"u}ltekin}, {Hazboun}, {Hourihane}, {Islo},
  {Jennings}, {Johnson}, {Jones}, {Kaiser}, {Kaplan}, {Kelley}, {Kerr}, {Key},
  {Laal}, {Lam}, {Lamb}, {Lazio}, {Lee}, {Lewandowska}, {Lino Dos Santos},
  {Littenberg}, {Liu}, {Lorimer}, {Luo}, {Lynch}, {Ma}, {Madison}, {McEwen},
  {McKee}, {McLaughlin}, {McMann}, {Meyers}, {Meyers}, {Mingarelli},
  {Mitridate}, {Nay}, {Natarajan}, {Ng}, {Nice}, {Ocker}, {Olum}, {Pennucci},
  {Perera}, {Petrov}, {Pol}, {Radovan}, {Ransom}, {Ray}, {Romano}, {Sardesai},
  {Schmiedekamp}, {Schmiedekamp}, {Schmitz}, {Schr{\"o}der}, {Schult},
  {Shapiro-Albert}, {Siemens}, {Simon}, {Siwek}, {Stairs}, {Stinebring},
  {Stovall}, {Stratmann}, {Sun}, {Susobhanan}, {Swiggum}, {Taylor}, {Taylor},
  {Trickle}, {Turner}, {Unal}, {Vallisneri}, {Verma}, {Vigeland}, {Wahl},
  {Wang}, {Witt}, {Wright}, {Young}, {Zurek}, \& {Nanograv
  Collaboration}}]{nanograv_15yr_new_physics}
{Afzal}, A., {Agazie}, G., {Anumarlapudi}, A., {et~al.} 2023, \apjl, 951, L11,
  \dodoi{10.3847/2041-8213/acdc91}

\bibitem[{{Agazie} {et~al.}(2023{\natexlab{a}}){Agazie}, {Anumarlapudi},
  {Archibald}, {Arzoumanian}, {Baker}, {B{\'e}csy}, {Blecha}, {Brazier},
  {Brook}, {Burke-Spolaor}, {Burnette}, {Case}, {Charisi}, {Chatterjee},
  {Chatziioannou}, {Cheeseboro}, {Chen}, {Cohen}, {Cordes}, {Cornish},
  {Crawford}, {Cromartie}, {Crowter}, {Cutler}, {Decesar}, {Degan}, {Demorest},
  {Deng}, {Dolch}, {Drachler}, {Ellis}, {Ferrara}, {Fiore}, {Fonseca},
  {Freedman}, {Garver-Daniels}, {Gentile}, {Gersbach}, {Glaser}, {Good},
  {G{\"u}ltekin}, {Hazboun}, {Hourihane}, {Islo}, {Jennings}, {Johnson},
  {Jones}, {Kaiser}, {Kaplan}, {Kelley}, {Kerr}, {Key}, {Klein}, {Laal}, {Lam},
  {Lamb}, {Lazio}, {Lewandowska}, {Littenberg}, {Liu}, {Lommen}, {Lorimer},
  {Luo}, {Lynch}, {Ma}, {Madison}, {Mattson}, {McEwen}, {McKee}, {McLaughlin},
  {McMann}, {Meyers}, {Meyers}, {Mingarelli}, {Mitridate}, {Natarajan}, {Ng},
  {Nice}, {Ocker}, {Olum}, {Pennucci}, {Perera}, {Petrov}, {Pol}, {Radovan},
  {Ransom}, {Ray}, {Romano}, {Sardesai}, {Schmiedekamp}, {Schmiedekamp},
  {Schmitz}, {Schult}, {Shapiro-Albert}, {Siemens}, {Simon}, {Siwek}, {Stairs},
  {Stinebring}, {Stovall}, {Sun}, {Susobhanan}, {Swiggum}, {Taylor}, {Taylor},
  {Turner}, {Unal}, {Vallisneri}, {van Haasteren}, {Vigeland}, {Wahl}, {Wang},
  {Witt}, {Young}, \& {Nanograv Collaboration}}]{nanograv_15yr_gwb}
{Agazie}, G., {Anumarlapudi}, A., {Archibald}, A.~M., {et~al.}
  2023{\natexlab{a}}, \apjl, 951, L8, \dodoi{10.3847/2041-8213/acdac6}

\bibitem[{{Agazie} {et~al.}(2023{\natexlab{b}}){Agazie}, {Anumarlapudi},
  {Archibald}, {Baker}, {B{\'e}csy}, {Blecha}, {Bonilla}, {Brazier}, {Brook},
  {Burke-Spolaor}, {Burnette}, {Case}, {Casey-Clyde}, {Charisi}, {Chatterjee},
  {Chatziioannou}, {Cheeseboro}, {Chen}, {Cohen}, {Cordes}, {Cornish},
  {Crawford}, {Cromartie}, {Crowter}, {Cutler}, {D'Orazio}, {Decesar}, {Degan},
  {Demorest}, {Deng}, {Dolch}, {Drachler}, {Ferrara}, {Fiore}, {Fonseca},
  {Freedman}, {Gardiner}, {Garver-Daniels}, {Gentile}, {Gersbach}, {Glaser},
  {Good}, {G{\"u}ltekin}, {Hazboun}, {Hourihane}, {Islo}, {Jennings},
  {Johnson}, {Jones}, {Kaiser}, {Kaplan}, {Kelley}, {Kerr}, {Key}, {Laal},
  {Lam}, {Lamb}, {Lazio}, {Lewandowska}, {Littenberg}, {Liu}, {Luo}, {Lynch},
  {Ma}, {Madison}, {McEwen}, {McKee}, {McLaughlin}, {McMann}, {Meyers},
  {Meyers}, {Mingarelli}, {Mitridate}, {Natarajan}, {Ng}, {Nice}, {Ocker},
  {Olum}, {Pennucci}, {Perera}, {Petrov}, {Pol}, {Radovan}, {Ransom}, {Ray},
  {Romano}, {Runnoe}, {Sardesai}, {Schmiedekamp}, {Schmiedekamp}, {Schmitz},
  {Schult}, {Shapiro-Albert}, {Siemens}, {Simon}, {Siwek}, {Stairs},
  {Stinebring}, {Stovall}, {Sun}, {Susobhanan}, {Swiggum}, {Taylor}, {Taylor},
  {Turner}, {Unal}, {Vallisneri}, {Vigeland}, {Wachter}, {Wahl}, {Wang},
  {Witt}, {Wright}, {Young}, \& {Nanograv Collaboration}}]{nanograv_15yr_astro}
---. 2023{\natexlab{b}}, \apjl, 952, L37, \dodoi{10.3847/2041-8213/ace18b}

\bibitem[{{Agazie} {et~al.}(2023{\natexlab{c}}){Agazie}, {Anumarlapudi},
  {Archibald}, {Arzoumanian}, {Baker}, {B{\'e}csy}, {Blecha}, {Brazier},
  {Brook}, {Burke-Spolaor}, {Charisi}, {Chatterjee}, {Cohen}, {Cordes},
  {Cornish}, {Crawford}, {Cromartie}, {Crowter}, {Decesar}, {Demorest},
  {Dolch}, {Drachler}, {Ferrara}, {Fiore}, {Fonseca}, {Freedman},
  {Garver-Daniels}, {Gentile}, {Glaser}, {Good}, {Guertin}, {G{\"u}ltekin},
  {Hazboun}, {Jennings}, {Johnson}, {Jones}, {Kaiser}, {Kaplan}, {Kelley},
  {Kerr}, {Key}, {Laal}, {Lam}, {Lamb}, {Lazio}, {Lewandowska}, {Liu},
  {Lorimer}, {Luo}, {Lynch}, {Ma}, {Madison}, {McEwen}, {McKee}, {McLaughlin},
  {McMann}, {Meyers}, {Mingarelli}, {Mitridate}, {Ng}, {Nice}, {Ocker}, {Olum},
  {Pennucci}, {Perera}, {Pol}, {Radovan}, {Ransom}, {Ray}, {Romano},
  {Sardesai}, {Schmiedekamp}, {Schmiedekamp}, {Schmitz}, {Shapiro-Albert},
  {Siemens}, {Simon}, {Siwek}, {Stairs}, {Stinebring}, {Stovall}, {Susobhanan},
  {Swiggum}, {Taylor}, {Turner}, {Unal}, {Vallisneri}, {Vigeland}, {Wahl},
  {Witt}, {Young}, \& {Nanograv Collaboration}}]{nanograv_15yr_detchar}
---. 2023{\natexlab{c}}, \apjl, 951, L10, \dodoi{10.3847/2041-8213/acda88}

\bibitem[{{Agazie} {et~al.}(2023{\natexlab{d}}){Agazie}, {Anumarlapudi},
  {Archibald}, {Arzoumanian}, {Baker}, {B{\'e}csy}, {Blecha}, {Brazier},
  {Brook}, {Burke-Spolaor}, {Casey-Clyde}, {Charisi}, {Chatterjee}, {Cohen},
  {Cordes}, {Cornish}, {Crawford}, {Cromartie}, {Crowter}, {DeCesar},
  {Demorest}, {Dolch}, {Drachler}, {Ferrara}, {Fiore}, {Fonseca}, {Freedman},
  {Gardiner}, {Garver-Daniels}, {Gentile}, {Glaser}, {Good}, {G{\"u}ltekin},
  {Hazboun}, {Jennings}, {Johnson}, {Jones}, {Kaiser}, {Kaplan}, {Kelley},
  {Kerr}, {Key}, {Laal}, {Lam}, {Lamb}, {Lazio}, {Lewandowska}, {Liu},
  {Lorimer}, {Luo}, {Lynch}, {Ma}, {Madison}, {McEwen}, {McKee}, {McLaughlin},
  {McMann}, {Meyers}, {Mingarelli}, {Mitridate}, {Ng}, {Nice}, {Ocker}, {Olum},
  {Pennucci}, {Perera}, {Pol}, {Radovan}, {Ransom}, {Ray}, {Romano},
  {Sardesai}, {Schmiedekamp}, {Schmiedekamp}, {Schmitz}, {Schult},
  {Shapiro-Albert}, {Siemens}, {Simon}, {Siwek}, {Stairs}, {Stinebring},
  {Stovall}, {Susobhanan}, {Swiggum}, {Taylor}, {Turner}, {Unal}, {Vallisneri},
  {Vigeland}, {Wahl}, {Witt}, \& {Young}}]{nanograv_15yr_anisotropy}
---. 2023{\natexlab{d}}, \apjl, 956, L3, \dodoi{10.3847/2041-8213/acf4fd}

\bibitem[{{Agazie} {et~al.}(2023{\natexlab{e}}){Agazie}, {Alam},
  {Anumarlapudi}, {Archibald}, {Arzoumanian}, {Baker}, {Blecha}, {Bonidie},
  {Brazier}, {Brook}, {Burke-Spolaor}, {B{\'e}csy}, {Chapman}, {Charisi},
  {Chatterjee}, {Cohen}, {Cordes}, {Cornish}, {Crawford}, {Cromartie},
  {Crowter}, {Decesar}, {Demorest}, {Dolch}, {Drachler}, {Ferrara}, {Fiore},
  {Fonseca}, {Freedman}, {Garver-Daniels}, {Gentile}, {Glaser}, {Good},
  {G{\"u}ltekin}, {Hazboun}, {Jennings}, {Jessup}, {Johnson}, {Jones},
  {Kaiser}, {Kaplan}, {Kelley}, {Kerr}, {Key}, {Kuske}, {Laal}, {Lam}, {Lamb},
  {Lazio}, {Lewandowska}, {Lin}, {Liu}, {Lorimer}, {Luo}, {Lynch}, {Ma},
  {Madison}, {Maraccini}, {McEwen}, {McKee}, {McLaughlin}, {McMann}, {Meyers},
  {Mingarelli}, {Mitridate}, {Ng}, {Nice}, {Ocker}, {Olum}, {Panciu},
  {Pennucci}, {Perera}, {Pol}, {Radovan}, {Ransom}, {Ray}, {Romano}, {Salo},
  {Sardesai}, {Schmiedekamp}, {Schmiedekamp}, {Schmitz}, {Shapiro-Albert},
  {Siemens}, {Simon}, {Siwek}, {Stairs}, {Stinebring}, {Stovall}, {Susobhanan},
  {Swiggum}, {Taylor}, {Turner}, {Unal}, {Vallisneri}, {Vigeland}, {Wahl},
  {Wang}, {Witt}, {Young}, \& {Nanograv Collaboration}}]{nanograv_15yr_dataset}
{Agazie}, G., {Alam}, M.~F., {Anumarlapudi}, A., {et~al.} 2023{\natexlab{e}},
  \apjl, 951, L9, \dodoi{10.3847/2041-8213/acda9a}

\bibitem[{{Agazie} {et~al.}(2023{\natexlab{f}}){Agazie}, {Anumarlapudi},
  {Archibald}, {Arzoumanian}, {Baker}, {B{\'e}csy}, {Blecha}, {Brazier},
  {Brook}, {Burke-Spolaor}, {Case}, {Casey-Clyde}, {Charisi}, {Chatterjee},
  {Cohen}, {Cordes}, {Cornish}, {Crawford}, {Cromartie}, {Crowter}, {Decesar},
  {Demorest}, {Digman}, {Dolch}, {Drachler}, {Ferrara}, {Fiore}, {Fonseca},
  {Freedman}, {Garver-Daniels}, {Gentile}, {Glaser}, {Good}, {G{\"u}ltekin},
  {Hazboun}, {Hourihane}, {Jennings}, {Johnson}, {Jones}, {Kaiser}, {Kaplan},
  {Kelley}, {Kerr}, {Key}, {Laal}, {Lam}, {Lamb}, {Lazio}, {Lewandowska},
  {Liu}, {Lorimer}, {Luo}, {Lynch}, {Ma}, {Madison}, {McEwen}, {McKee},
  {McLaughlin}, {McMann}, {Meyers}, {Meyers}, {Mingarelli}, {Mitridate}, {Ng},
  {Nice}, {Ocker}, {Olum}, {Pennucci}, {Perera}, {Petrov}, {Pol}, {Radovan},
  {Ransom}, {Ray}, {Romano}, {Sardesai}, {Schmiedekamp}, {Schmiedekamp},
  {Schmitz}, {Shapiro-Albert}, {Siemens}, {Simon}, {Siwek}, {Stairs},
  {Stinebring}, {Stovall}, {Susobhanan}, {Swiggum}, {Taylor}, {Taylor},
  {Turner}, {Unal}, {Vallisneri}, {van Haasteren}, {Vigeland}, {Wahl}, {Witt},
  {Young}, \& {Nanograv Collaboration}}]{nanograv_15yr_cw}
{Agazie}, G., {Anumarlapudi}, A., {Archibald}, A.~M., {et~al.}
  2023{\natexlab{f}}, \apjl, 951, L50, \dodoi{10.3847/2041-8213/ace18a}

\bibitem[{{Ali-Ha{\"\i}moud} {et~al.}(2020){Ali-Ha{\"\i}moud}, {Smith}, \&
  {Mingarelli}}]{Chiara_anisotropy_2020}
{Ali-Ha{\"\i}moud}, Y., {Smith}, T.~L., \& {Mingarelli}, C. M.~F. 2020, \prd,
  102, 122005, \dodoi{10.1103/PhysRevD.102.122005}

\bibitem[{{Antoniadis} {et~al.}(2023{\natexlab{a}}){Antoniadis}, {Arumugam},
  {Arumugam}, {Babak}, {Bagchi}, {Bak Nielsen}, {Bassa}, {Bathula},
  {Berthereau}, {Bonetti}, {Bortolas}, {Brook}, {Burgay}, {Caballero},
  {Chalumeau}, {Champion}, {Chanlaridis}, {Chen}, {Cognard}, {Dandapat}, {Deb},
  {Desai}, {Desvignes}, {Dhanda-Batra}, {Dwivedi}, {Falxa}, {Ferdman},
  {Franchini}, {Gair}, {Goncharov}, {Gopakumar}, {Graikou}, {Grie{\ss}meier},
  {Guillemot}, {Guo}, {Gupta}, {Hisano}, {Hu}, {Iraci}, {Izquierdo-Villalba},
  {Jang}, {Jawor}, {Janssen}, {Jessner}, {Joshi}, {Kareem}, {Karuppusamy},
  {Keane}, {Keith}, {Kharbanda}, {Kikunaga}, {Kolhe}, {Kramer}, {Krishnakumar},
  {Lackeos}, {Lee}, {Liu}, {Liu}, {Lyne}, {McKee}, {Maan}, {Main},
  {Mickaliger}, {Ni{\c{t}}u}, {Nobleson}, {Paladi}, {Parthasarathy}, {Perera},
  {Perrodin}, {Petiteau}, {Porayko}, {Possenti}, {Prabu}, {Quelquejay Leclere},
  {Rana}, {Samajdar}, {Sanidas}, {Sesana}, {Shaifullah}, {Singha}, {Speri},
  {Spiewak}, {Srivastava}, {Stappers}, {Surnis}, {Susarla}, {Susobhanan},
  {Takahashi}, {Tarafdar}, {Theureau}, {Tiburzi}, {van der Wateren}, {Vecchio},
  {Venkatraman Krishnan}, {Verbiest}, {Wang}, {Wang}, \& {Wu}}]{epta_dr2_gwb}
{Antoniadis}, J., {Arumugam}, P., {Arumugam}, S., {et~al.} 2023{\natexlab{a}},
  \aap, 678, A50, \dodoi{10.1051/0004-6361/202346844}

\bibitem[{{Antoniadis} {et~al.}(2023{\natexlab{b}}){Antoniadis}, {Arumugam},
  {Arumugam}, {Babak}, {Bagchi}, {Bak Nielsen}, {Bassa}, {Bathula},
  {Berthereau}, {Bonetti}, {Bortolas}, {Brook}, {Burgay}, {Caballero},
  {Chalumeau}, {Champion}, {Chanlaridis}, {Chen}, {Cognard}, {Dandapat}, {Deb},
  {Desai}, {Desvignes}, {Dhanda-Batra}, {Dwivedi}, {Falxa}, {Ferranti},
  {Ferdman}, {Franchini}, {Gair}, {Goncharov}, {Gopakumar}, {Graikou},
  {Grie{\ss}meier}, {Guillemot}, {Guo}, {Gupta}, {Hisano}, {Hu}, {Iraci},
  {Izquierdo-Villalba}, {Jang}, {Jawor}, {Janssen}, {Jessner}, {Joshi},
  {Kareem}, {Karuppusamy}, {Keane}, {Keith}, {Kharbanda}, {Kikunaga}, {Kolhe},
  {Kramer}, {Krishnakumar}, {Lackeos}, {Lee}, {Liu}, {Liu}, {Lyne}, {McKee},
  {Maan}, {Main}, {Manzini}, {Mickaliger}, {Nitu}, {Nobleson}, {Paladi},
  {Parthasarathy}, {Perera}, {Perrodin}, {Petiteau}, {Porayko}, {Possenti},
  {Prabu}, {Quelquejay Leclere}, {Rana}, {Samajdar}, {Sanidas}, {Sesana},
  {Shaifullah}, {Singha}, {Speri}, {Spiewak}, {Srivastava}, {Stappers},
  {Surnis}, {Susarla}, {Susobhanan}, {Takahashi}, {Tarafdar}, {Theureau},
  {Tiburzi}, {van der Wateren}, {Vecchio}, {Venkatraman Krishnan}, {Verbiest},
  {Wang}, {Wang}, \& {Wu}}]{epta_dr2_cw}
---. 2023{\natexlab{b}}, arXiv e-prints, arXiv:2306.16226,
  \dodoi{10.48550/arXiv.2306.16226}

\bibitem[{{Astropy Collaboration} {et~al.}(2013){Astropy Collaboration},
  {Robitaille}, {Tollerud}, {Greenfield}, {Droettboom}, {Bray}, {Aldcroft},
  {Davis}, {Ginsburg}, {Price-Whelan}, {Kerzendorf}, {Conley}, {Crighton},
  {Barbary}, {Muna}, {Ferguson}, {Grollier}, {Parikh}, {Nair}, {Unther},
  {Deil}, {Woillez}, {Conseil}, {Kramer}, {Turner}, {Singer}, {Fox}, {Weaver},
  {Zabalza}, {Edwards}, {Azalee Bostroem}, {Burke}, {Casey}, {Crawford},
  {Dencheva}, {Ely}, {Jenness}, {Labrie}, {Lim}, {Pierfederici}, {Pontzen},
  {Ptak}, {Refsdal}, {Servillat}, \& {Streicher}}]{astropy:2013}
{Astropy Collaboration}, {Robitaille}, T.~P., {Tollerud}, E.~J., {et~al.} 2013,
  \aap, 558, A33, \dodoi{10.1051/0004-6361/201322068}

\bibitem[{{B{\'e}csy} \& {Cornish}(2020)}]{BayesHopper}
{B{\'e}csy}, B., \& {Cornish}, N.~J. 2020, Classical and Quantum Gravity, 37,
  135011, \dodoi{10.1088/1361-6382/ab8bbd}

\bibitem[{{B{\'e}csy} {et~al.}(2022){B{\'e}csy}, {Cornish}, \&
  {Digman}}]{QuickCW}
{B{\'e}csy}, B., {Cornish}, N.~J., \& {Digman}, M.~C. 2022, \prd, 105, 122003,
  \dodoi{10.1103/PhysRevD.105.122003}

\bibitem[{B\'ecsy {et~al.}(2022)B\'ecsy, Cornish, \& Kelley}]{Becsy:2022pnr}
B\'ecsy, B., Cornish, N.~J., \& Kelley, L.~Z. 2022, Astrophys. J., 941, 119,
  \dodoi{10.3847/1538-4357/aca1b2}

\bibitem[{B{\'e}csy {et~al.}(2023)B{\'e}csy, Digman, \& Cornish}]{QuickCW_code}
B{\'e}csy, B., Digman, M., \& Cornish, N.~J. 2023, QuickCW (v1.0.1).
\newblock \url{https://github.com/nanograv/QuickCW/tree/v1.0.1}

\bibitem[{{Chamberlin} {et~al.}(2015){Chamberlin}, {Creighton}, {Siemens},
  {Demorest}, {Ellis}, {Price}, \& {Romano}}]{OS}
{Chamberlin}, S.~J., {Creighton}, J. D.~E., {Siemens}, X., {et~al.} 2015, \prd,
  91, 044048, \dodoi{10.1103/PhysRevD.91.044048}

\bibitem[{{Cornish} \& {Sampson}(2016)}]{cornish_sampson_2016}
{Cornish}, N.~J., \& {Sampson}, L. 2016, \prd, 93, 104047,
  \dodoi{10.1103/PhysRevD.93.104047}

\bibitem[{Cornish \& Sesana(2013)}]{Cornish:2013aba}
Cornish, N.~J., \& Sesana, A. 2013, Class. Quant. Grav., 30, 224005,
  \dodoi{10.1088/0264-9381/30/22/224005}

\bibitem[{{Ellis} {et~al.}(2019){Ellis}, {Vallisneri}, {Taylor}, \&
  {Baker}}]{enterprise}
{Ellis}, J.~A., {Vallisneri}, M., {Taylor}, S.~R., \& {Baker}, P.~T. 2019,
  {ENTERPRISE: Enhanced Numerical Toolbox Enabling a Robust PulsaR Inference
  SuitE}.
\newblock \doeprint{1912.015}

\bibitem[{Foreman-Mackey(2016)}]{corner}
Foreman-Mackey, D. 2016, The Journal of Open Source Software, 1, 24,
  \dodoi{10.21105/joss.00024}

\bibitem[{{Gair} {et~al.}(2015){Gair}, {Romano}, \&
  {Taylor}}]{Gair_anisotropy_2015}
{Gair}, J.~R., {Romano}, J.~D., \& {Taylor}, S.~R. 2015, \prd, 92, 102003,
  \dodoi{10.1103/PhysRevD.92.102003}

\bibitem[{{Gardiner} {et~al.}(2023){Gardiner}, {Kelley}, {Lemke}, \&
  {Mitridate}}]{Emiko_cw_paper}
{Gardiner}, E.~C., {Kelley}, L.~Z., {Lemke}, A.-M., \& {Mitridate}, A. 2023,
  arXiv e-prints, arXiv:2309.07227, \dodoi{10.48550/arXiv.2309.07227}

\bibitem[{{Gersbach} {et~al.}(2023){Gersbach}, {Meyers}, {Taylor}, {Joseph},
  {et~al.}}]{free_spectrum_os}
{Gersbach}, K., {Meyers}, P., {Taylor}, S., {Joseph}, R., {et~al.} 2023, in
  preparation

\bibitem[{{Goncharov} {et~al.}(2021){Goncharov}, {Shannon}, {Reardon}, {Hobbs},
  {Zic}, {Bailes}, {Cury{\l}o}, {Dai}, {Kerr}, {Lower}, {Manchester}, {Mandow},
  {Middleton}, {Miles}, {Parthasarathy}, {Thrane}, {Thyagarajan}, {Xue}, {Zhu},
  {Cameron}, {Feng}, {Luo}, {Russell}, {Sarkissian}, {Spiewak}, {Wang}, {Wang},
  {Zhang}, \& {Zhang}}]{goncharov2021}
{Goncharov}, B., {Shannon}, R.~M., {Reardon}, D.~J., {et~al.} 2021, \apjl, 917,
  L19, \dodoi{10.3847/2041-8213/ac17f4}

\bibitem[{{Hellings} \& {Downs}(1983)}]{HD}
{Hellings}, R.~W., \& {Downs}, G.~S. 1983, \apjl, 265, L39,
  \dodoi{10.1086/183954}

\bibitem[{{Hobbs} {et~al.}(2006){Hobbs}, {Edwards}, \& {Manchester}}]{tempo2}
{Hobbs}, G.~B., {Edwards}, R.~T., \& {Manchester}, R.~N. 2006, \mnras, 369,
  655, \dodoi{10.1111/j.1365-2966.2006.10302.x}

\bibitem[{{Hourihane} {et~al.}(2023){Hourihane}, {Meyers}, {Johnson},
  {Chatziioannou}, \& {Vallisneri}}]{sophie_resampling}
{Hourihane}, S., {Meyers}, P., {Johnson}, A., {Chatziioannou}, K., \&
  {Vallisneri}, M. 2023, \prd, 107, 084045, \dodoi{10.1103/PhysRevD.107.084045}

\bibitem[{Hunter(2007)}]{matplotlib}
Hunter, J.~D. 2007, Computing in Science \& Engineering, 9, 90,
  \dodoi{10.1109/MCSE.2007.55}

\bibitem[{{Johnson} {et~al.}(2023){Johnson}, {Meyers}, {Baker}, {Cornish},
  {Hazboun}, {Littenberg}, {Romano}, {Taylor}, {Vallisneri}, {Vigeland},
  {Olum}, {Siemens}, {Ellis}, {van Haasteren}, {Hourihane}, {Agazie},
  {Anumarlapudi}, {Archibald}, {Arzoumanian}, {Blecha}, {Brazier}, {Brook},
  {Burke-Spolaor}, {B{\'e}csy}, {Casey-Clyde}, {Charisi}, {Chatterjee},
  {Chatziioannou}, {Cohen}, {Cordes}, {Crawford}, {Cromartie}, {Crowter},
  {DeCesar}, {Demorest}, {Dolch}, {Drachler}, {Ferrara}, {Fiore}, {Fonseca},
  {Freedman}, {Garver-Daniels}, {Gentile}, {Glaser}, {Good}, {G{\"u}ltekin},
  {Jennings}, {Jones}, {Kaiser}, {Kaplan}, {Kelley}, {Kerr}, {Key}, {Laal},
  {Lam}, {Lamb}, {Lazio}, {Lewandowska}, {Liu}, {Lorimer}, {Lynch}, {Ma},
  {Madison}, {McEwen}, {McKee}, {McLaughlin}, {McMann}, {Meyers}, {Mingarelli},
  {Mitridate}, {Ng}, {Nice}, {Ocker}, {Pennucci}, {Perera}, {Pol}, {Radovan},
  {Ransom}, {Ray}, {Sardesai}, {Schmiedekamp}, {Schmiedekamp}, {Schmitz},
  {Shapiro-Albert}, {Simon}, {Siwek}, {Stairs}, {Stinebring}, {Stovall},
  {Susobhanan}, {Swiggum}, {Turner}, {Unal}, {Wahl}, {Witt}, \&
  {Young}}]{nanograv_15yr_code_review}
{Johnson}, A.~D., {Meyers}, P.~M., {Baker}, P.~T., {et~al.} 2023, arXiv
  e-prints, arXiv:2306.16223, \dodoi{10.48550/arXiv.2306.16223}

\bibitem[{{Kelley} {et~al.}(2017{\natexlab{a}}){Kelley}, {Blecha}, \&
  {Hernquist}}]{Luke_paper1}
{Kelley}, L.~Z., {Blecha}, L., \& {Hernquist}, L. 2017{\natexlab{a}}, \mnras,
  464, 3131, \dodoi{10.1093/mnras/stw2452}

\bibitem[{{Kelley} {et~al.}(2017{\natexlab{b}}){Kelley}, {Blecha}, {Hernquist},
  {Sesana}, \& {Taylor}}]{Luke_paper2}
{Kelley}, L.~Z., {Blecha}, L., {Hernquist}, L., {Sesana}, A., \& {Taylor},
  S.~R. 2017{\natexlab{b}}, \mnras, 471, 4508, \dodoi{10.1093/mnras/stx1638}

\bibitem[{{Kelley} {et~al.}(2018){Kelley}, {Blecha}, {Hernquist}, {Sesana}, \&
  {Taylor}}]{Luke_single_source}
---. 2018, \mnras, 477, 964, \dodoi{10.1093/mnras/sty689}

\bibitem[{{Kelley} {et~al.}(2023)}]{holodeck}
{Kelley}, L.~Z., {et~al.} 2023, in preparation

\bibitem[{{Lamb} {et~al.}(2023){Lamb}, {Taylor}, \& {van
  Haasteren}}]{William_spectral_refitting}
{Lamb}, W.~G., {Taylor}, S.~R., \& {van Haasteren}, R. 2023, \prd, 108, 103019,
  \dodoi{10.1103/PhysRevD.108.103019}

\bibitem[{{Lentati} {et~al.}(2013){Lentati}, {Alexander}, {Hobson}, {Taylor},
  {Gair}, {Balan}, \& {van Haasteren}}]{lentati_et_al_2013}
{Lentati}, L., {Alexander}, P., {Hobson}, M.~P., {et~al.} 2013, \prd, 87,
  104021, \dodoi{10.1103/PhysRevD.87.104021}

\bibitem[{{Luo} {et~al.}(2019){Luo}, {Ransom}, {Demorest}, {van Haasteren},
  {Ray}, {Stovall}, {Bachetti}, {Archibald}, {Kerr}, {Colen}, \&
  {Jenet}}]{pint}
{Luo}, J., {Ransom}, S., {Demorest}, P., {et~al.} 2019, {PINT: High-precision
  pulsar timing analysis package}, Astrophysics Source Code Library, record
  ascl:1902.007.
\newblock \doeprint{1902.007}

\bibitem[{{Meyers} {et~al.}(2023){Meyers}, {Chatziioannou}, {Vallisneri}, \&
  {Chua}}]{Pat_post_Pred_check}
{Meyers}, P.~M., {Chatziioannou}, K., {Vallisneri}, M., \& {Chua}, A. J.~K.
  2023, arXiv e-prints, arXiv:2306.05559, \dodoi{10.48550/arXiv.2306.05559}

\bibitem[{{Mingarelli} {et~al.}(2013){Mingarelli}, {Sidery}, {Mandel}, \&
  {Vecchio}}]{Chiara_anisotropy_2013}
{Mingarelli}, C.~M.~F., {Sidery}, T., {Mandel}, I., \& {Vecchio}, A. 2013,
  \prd, 88, 062005, \dodoi{10.1103/PhysRevD.88.062005}

\bibitem[{{Nice} {et~al.}(2015){Nice}, {Demorest}, {Stairs}, {Manchester},
  {Taylor}, {Peters}, {Weisberg}, {Irwin}, {Wex}, \& {Huang}}]{tempo}
{Nice}, D., {Demorest}, P., {Stairs}, I., {et~al.} 2015, {Tempo: Pulsar timing
  data analysis}, Astrophysics Source Code Library, record ascl:1509.002.
\newblock \doeprint{1509.002}

\bibitem[{{Pol} {et~al.}(2022){Pol}, {Taylor}, \&
  {Romano}}]{Nihan_anisotropy_forecast}
{Pol}, N., {Taylor}, S.~R., \& {Romano}, J.~D. 2022, \apj, 940, 173,
  \dodoi{10.3847/1538-4357/ac9836}

\bibitem[{{Pol} {et~al.}(2021){Pol}, {Taylor}, {Kelley}, {Vigeland}, {Simon},
  {Chen}, {Arzoumanian}, {Baker}, {B{\'e}csy}, {Brazier}, {Brook},
  {Burke-Spolaor}, {Chatterjee}, {Cordes}, {Cornish}, {Crawford}, {Thankful
  Cromartie}, {Decesar}, {Demorest}, {Dolch}, {Ferrara}, {Fiore}, {Fonseca},
  {Garver-Daniels}, {Good}, {Hazboun}, {Jennings}, {Jones}, {Kaiser}, {Kaplan},
  {Shapiro Key}, {Lam}, {Lazio}, {Luo}, {Lynch}, {Madison}, {McEwen},
  {McLaughlin}, {Mingarelli}, {Ng}, {Nice}, {Pennucci}, {Ransom}, {Ray},
  {Shapiro-Albert}, {Siemens}, {Stairs}, {Stinebring}, {Swiggum}, {Vallisneri},
  {Wahl}, {Witt}, \& {Nanograv Collaboration}}]{astro4cast}
{Pol}, N.~S., {Taylor}, S.~R., {Kelley}, L.~Z., {et~al.} 2021, \apjl, 911, L34,
  \dodoi{10.3847/2041-8213/abf2c9}

\bibitem[{{Price-Whelan} {et~al.}(2018){Price-Whelan}, {Sip{\H{o}}cz},
  {G{\"u}nther}, {Lim}, {Crawford}, {Conseil}, {Shupe}, {Craig}, {Dencheva},
  {Ginsburg}, {VanderPlas}, {Bradley}, {P{\'e}rez-Su{\'a}rez}, {de Val-Borro},
  {Paper Contributors}, {Aldcroft}, {Cruz}, {Robitaille}, {Tollerud},
  {Coordination Committee}, {Ardelean}, {Babej}, {Bach}, {Bachetti}, {Bakanov},
  {Bamford}, {Barentsen}, {Barmby}, {Baumbach}, {Berry}, {Biscani}, {Boquien},
  {Bostroem}, {Bouma}, {Brammer}, {Bray}, {Breytenbach}, {Buddelmeijer},
  {Burke}, {Calderone}, {Cano Rodr{\'\i}guez}, {Cara}, {Cardoso}, {Cheedella},
  {Copin}, {Corrales}, {Crichton}, {D{\textquoteright}Avella}, {Deil},
  {Depagne}, {Dietrich}, {Donath}, {Droettboom}, {Earl}, {Erben}, {Fabbro},
  {Ferreira}, {Finethy}, {Fox}, {Garrison}, {Gibbons}, {Goldstein}, {Gommers},
  {Greco}, {Greenfield}, {Groener}, {Grollier}, {Hagen}, {Hirst}, {Homeier},
  {Horton}, {Hosseinzadeh}, {Hu}, {Hunkeler}, {Ivezi{\'c}}, {Jain}, {Jenness},
  {Kanarek}, {Kendrew}, {Kern}, {Kerzendorf}, {Khvalko}, {King}, {Kirkby},
  {Kulkarni}, {Kumar}, {Lee}, {Lenz}, {Littlefair}, {Ma}, {Macleod},
  {Mastropietro}, {McCully}, {Montagnac}, {Morris}, {Mueller}, {Mumford},
  {Muna}, {Murphy}, {Nelson}, {Nguyen}, {Ninan}, {N{\"o}the}, {Ogaz}, {Oh},
  {Parejko}, {Parley}, {Pascual}, {Patil}, {Patil}, {Plunkett}, {Prochaska},
  {Rastogi}, {Reddy Janga}, {Sabater}, {Sakurikar}, {Seifert}, {Sherbert},
  {Sherwood-Taylor}, {Shih}, {Sick}, {Silbiger}, {Singanamalla}, {Singer},
  {Sladen}, {Sooley}, {Sornarajah}, {Streicher}, {Teuben}, {Thomas},
  {Tremblay}, {Turner}, {Terr{\'o}n}, {van Kerkwijk}, {de la Vega}, {Watkins},
  {Weaver}, {Whitmore}, {Woillez}, {Zabalza}, \& {Contributors}}]{astropy}
{Price-Whelan}, A.~M., {Sip{\H{o}}cz}, B.~M., {G{\"u}nther}, H.~M., {et~al.}
  2018, \aj, 156, 123, \dodoi{10.3847/1538-3881/aabc4f}

\bibitem[{{Reardon} {et~al.}(2023){Reardon}, {Zic}, {Shannon}, {Hobbs},
  {Bailes}, {Di Marco}, {Kapur}, {Rogers}, {Thrane}, {Askew}, {Bhat},
  {Cameron}, {Cury{\l}o}, {Coles}, {Dai}, {Goncharov}, {Kerr}, {Kulkarni},
  {Levin}, {Lower}, {Manchester}, {Mandow}, {Miles}, {Nathan}, {Os{\l}owski},
  {Russell}, {Spiewak}, {Zhang}, \& {Zhu}}]{ppta_dr3_gwb}
{Reardon}, D.~J., {Zic}, A., {Shannon}, R.~M., {et~al.} 2023, \apjl, 951, L6,
  \dodoi{10.3847/2041-8213/acdd02}

\bibitem[{{Romano} {et~al.}(2015){Romano}, {Taylor}, {Cornish}, {Gair},
  {Mingarelli}, \& {van Haasteren}}]{Romano_anisotropy_2015}
{Romano}, J.~D., {Taylor}, S.~R., {Cornish}, N.~J., {et~al.} 2015, \prd, 92,
  042003, \dodoi{10.1103/PhysRevD.92.042003}

\bibitem[{Sampson {et~al.}(2015)Sampson, Cornish, \&
  McWilliams}]{Sampson:2015ada}
Sampson, L., Cornish, N.~J., \& McWilliams, S.~T. 2015, Phys. Rev. D, 91,
  084055, \dodoi{10.1103/PhysRevD.91.084055}

\bibitem[{{Sardesai} \& {Vigeland}(2023)}]{mcos}
{Sardesai}, S.~C., \& {Vigeland}, S.~J. 2023, arXiv e-prints, arXiv:2303.09615,
  \dodoi{10.48550/arXiv.2303.09615}

\bibitem[{{Taylor}(2021)}]{steve_book}
{Taylor}, S.~R. 2021, arXiv e-prints, arXiv:2105.13270,
  \dodoi{10.48550/arXiv.2105.13270}

\bibitem[{Taylor {et~al.}(2021)Taylor, Baker, Hazboun, Simon, \& Vigeland}]{ee}
Taylor, S.~R., Baker, P.~T., Hazboun, J.~S., Simon, J., \& Vigeland, S.~J.
  2021, enterprise\textunderscore extensions.
\newblock \url{https://github.com/nanograv/enterprise_extensions}

\bibitem[{{Taylor} \& {Gair}(2013)}]{Steve_anisotropy_2013}
{Taylor}, S.~R., \& {Gair}, J.~R. 2013, \prd, 88, 084001,
  \dodoi{10.1103/PhysRevD.88.084001}

\bibitem[{{Taylor} {et~al.}(2017){Taylor}, {Simon}, \&
  {Sampson}}]{Steve_astro_spectral_modeling}
{Taylor}, S.~R., {Simon}, J., \& {Sampson}, L. 2017, \prl, 118, 181102,
  \dodoi{10.1103/PhysRevLett.118.181102}

\bibitem[{{Taylor} {et~al.}(2020){Taylor}, {van Haasteren}, \&
  {Sesana}}]{Steve_bumpy_background}
{Taylor}, S.~R., {van Haasteren}, R., \& {Sesana}, A. 2020, \prd, 102, 084039,
  \dodoi{10.1103/PhysRevD.102.084039}

\bibitem[{{The International Pulsar Timing Array Collaboration}
  {et~al.}(2023){The International Pulsar Timing Array Collaboration},
  {Agazie}, {Antoniadis}, {Anumarlapudi}, {Archibald}, {Arumugam}, {Arumugam},
  {Arzoumanian}, {Askew}, {Babak}, {Bagchi}, {Bailes}, {Bak Nielsen}, {Baker},
  {Bassa}, {Bathula}, {B{\'e}csy}, {Berthereau}, {Bhat}, {Blecha}, {Bonetti},
  {Bortolas}, {Brazier}, {Brook}, {Burgay}, {Burke-Spolaor}, {Burnette},
  {Caballero}, {Cameron}, {Case}, {Chalumeau}, {Champion}, {Chanlaridis},
  {Charisi}, {Chatterjee}, {Chatziioannou}, {Cheeseboro}, {Chen}, {Chen},
  {Cognard}, {Cohen}, {Coles}, {Cordes}, {Cornish}, {Crawford}, {Cromartie},
  {Crowter}, {Cury{\l}o}, {Cutler}, {Dai}, {Dandapat}, {Deb}, {DeCesar},
  {DeGan}, {Demorest}, {Deng}, {Desai}, {Desvignes}, {Dey}, {Dhanda-Batra}, {Di
  Marco}, {Dolch}, {Drachler}, {Dwivedi}, {Ellis}, {Falxa}, {Feng}, {Ferdman},
  {Ferrara}, {Fiore}, {Fonseca}, {Franchini}, {Freedman}, {Gair},
  {Garver-Daniels}, {Gentile}, {Gersbach}, {Glaser}, {Good}, {Goncharov},
  {Gopakumar}, {Graikou}, {Grie{\ss}meier}, {Guillemot}, {G{\"u}ltekin}, {Guo},
  {Gupta}, {Grunthal}, {Hazboun}, {Hisano}, {Hobbs}, {Hourihane}, {Hu},
  {Iraci}, {Islo}, {Izquierdo-Villalba}, {Jang}, {Jawor}, {Janssen},
  {Jennings}, {Jessner}, {Johnson}, {Jones}, {Joshi}, {Kaiser}, {Kaplan},
  {Kapur}, {Kareem}, {Karuppusamy}, {Keane}, {Keith}, {Kelley}, {Kerr}, {Key},
  {Kharbanda}, {Kikunaga}, {Klein}, {Kolhe}, {Kramer}, {Krishnakumar},
  {Kulkarni}, {Laal}, {Lackeos}, {Lam}, {Lamb}, {Larsen}, {Lazio}, {Lee},
  {Levin}, {Lewandowska}, {Littenberg}, {Liu}, {Liu}, {Liu}, {Lommen},
  {Lorimer}, {Lower}, {Luo}, {Luo}, {Lynch}, {Lyne}, {Ma}, {Maan}, {Madison},
  {Main}, {Manchester}, {Mandow}, {Mattson}, {McEwen}, {McKee}, {McLaughlin},
  {McMann}, {Meyers}, {Meyers}, {Mickaliger}, {Miles}, {Mingarelli},
  {Mitridate}, {Natarajan}, {Nathan}, {Ng}, {Nice}, {Ni{\c{t}}u}, {Nobleson},
  {Ocker}, {Olum}, {Os{\l}owski}, {Paladi}, {Parthasarathy}, {Pennucci},
  {Perera}, {Perrodin}, {Petiteau}, {Petrov}, {Pol}, {Porayko}, {Possenti},
  {Prabu}, {Quelquejay Leclere}, {Radovan}, {Rana}, {Ransom}, {Ray}, {Reardon},
  {Rogers}, {Romano}, {Russell}, {Samajdar}, {Sanidas}, {Sardesai},
  {Schmiedekamp}, {Schmiedekamp}, {Schmitz}, {Schult}, {Sesana}, {Shaifullah},
  {Shannon}, {Shapiro-Albert}, {Siemens}, {Simon}, {Singha}, {Siwek}, {Speri},
  {Spiewak}, {Srivastava}, {Stairs}, {Stappers}, {Stinebring}, {Stovall},
  {Sun}, {Surnis}, {Susarla}, {Susobhanan}, {Swiggum}, {Takahashi}, {Tarafdar},
  {Taylor}, {Taylor}, {Theureau}, {Thrane}, {Thyagarajan}, {Tiburzi}, {Toomey},
  {Turner}, {Unal}, {Vallisneri}, {van der Wateren}, {van Haasteren},
  {Vecchio}, {Venkatraman Krishnan}, {Verbiest}, {Vigeland}, {Wahl}, {Wang},
  {Wang}, {Witt}, {Wang}, {Wang}, {Wayt}, {Wu}, {Young}, {Zhang}, {Zhang},
  {Zhu}, \& {Zic}}]{3p_plus_comparison}
{The International Pulsar Timing Array Collaboration}, {Agazie}, G.,
  {Antoniadis}, J., {et~al.} 2023, arXiv e-prints, arXiv:2309.00693,
  \dodoi{10.48550/arXiv.2309.00693}

\bibitem[{{Vallisneri}(2020)}]{libstempo}
{Vallisneri}, M. 2020, {libstempo: Python wrapper for Tempo2}, Astrophysics
  Source Code Library, record ascl:2002.017.
\newblock \doeprint{2002.017}

\bibitem[{{Vallisneri} {et~al.}(2023){Vallisneri}, {Meyers}, {Chatziioannou},
  \& {Chua}}]{Michele_post_pred_check}
{Vallisneri}, M., {Meyers}, P.~M., {Chatziioannou}, K., \& {Chua}, A. J.~K.
  2023, arXiv e-prints, arXiv:2306.05558, \dodoi{10.48550/arXiv.2306.05558}

\bibitem[{{Valtolina} {et~al.}(2023){Valtolina}, {Shaifullah}, {Samajdar}, \&
  {Sesana}}]{valtolina2023}
{Valtolina}, S., {Shaifullah}, G., {Samajdar}, A., \& {Sesana}, A. 2023, arXiv
  e-prints, arXiv:2309.13117, \dodoi{10.48550/arXiv.2309.13117}

\bibitem[{{van Haasteren} \& {Vallisneri}(2014)}]{rutger_michele_2014}
{van Haasteren}, R., \& {Vallisneri}, M. 2014, \prd, 90, 104012,
  \dodoi{10.1103/PhysRevD.90.104012}

\bibitem[{{Vigeland} {et~al.}(2018){Vigeland}, {Islo}, {Taylor}, \&
  {Ellis}}]{nmos}
{Vigeland}, S.~J., {Islo}, K., {Taylor}, S.~R., \& {Ellis}, J.~A. 2018, \prd,
  98, 044003, \dodoi{10.1103/PhysRevD.98.044003}

\bibitem[{{Vogelsberger} {et~al.}(2014){Vogelsberger}, {Genel}, {Springel},
  {Torrey}, {Sijacki}, {Xu}, {Snyder}, {Nelson}, \& {Hernquist}}]{Illustris}
{Vogelsberger}, M., {Genel}, S., {Springel}, V., {et~al.} 2014, \mnras, 444,
  1518, \dodoi{10.1093/mnras/stu1536}

\bibitem[{{Xu} {et~al.}(2023){Xu}, {Chen}, {Guo}, {Jiang}, {Wang}, {Xu}, {Xue},
  {Nicolas Caballero}, {Yuan}, {Xu}, {Wang}, {Hao}, {Luo}, {Lee}, {Han},
  {Jiang}, {Shen}, {Wang}, {Wang}, {Xu}, {Wu}, {Manchester}, {Qian}, {Guan},
  {Huang}, {Sun}, \& {Zhu}}]{cpta_gwb}
{Xu}, H., {Chen}, S., {Guo}, Y., {et~al.} 2023, Research in Astronomy and
  Astrophysics, 23, 075024, \dodoi{10.1088/1674-4527/acdfa5}

\end{thebibliography}
\bibliographystyle{aasjournal}



\end{document}